\documentstyle[aaspp4,epsfig,natbib209]{article}
\def\simlt{\lower.5ex\hbox{$\; \buildrel < \over \sim \;$}}
\def\simgt{\lower.5ex\hbox{$\; \buildrel > \over \sim \;$}}

\def\kms{\mbox{ km s$^{-1}$}}

\def\kpc{\mbox{ kpc}}

\def\msun{\mbox{ M}_\odot}

\begin{document}

\title{Simulations of Strong Gravitational Lensing with Substructure}

\author{Adam Amara}
\affil{\it Institute of Astronomy, Cambridge University, Madingley Road,
  CB3 0HA United Kingdom}
\and
\author{R. Benton Metcalf\footnote{Hubble Fellow}}
\affil{\it Department of Astronomy and Astrophysics, University of California,
Santa Cruz, CA 95064 USA}
\and
\author{Thomas J. Cox}
\affil{\it Department of Physics, University of California,
Santa Cruz, CA 95064 USA}
\and
\author{Jeremiah P. Ostriker}
\affil{\it Institute of Astronomy, Cambridge University, Madingley Road,
  CB3 0HA United Kingdom}

\abstract{
Galactic sized gravitational lenses are simulated by combining a 
cosmological N-body simulation and models for the baryonic component of
the galaxy.  The lens caustics, critical curves, image locations and
magnification ratios are calculated by ray-shooting on an adaptive grid.
When the source is near a cusp in a smooth lens' caustic the sum of the
magnifications of the three closest images should be close to zero.
It is found that in the observed cases this sum is generally too large
to be consistent with the simulations implying that there is not
enough substructure in the simulations.
This suggests that other factors play an important role.  These may
include limited numerical resolution, lensing by structure outside the
halo, selection bias and the possibility that a randomly selected
galaxy halo may be more irregular, for example due to recent mergers,
than the isolated halo used in this study.  It is also shown that,
with the level of substructure computed from the N-body simulations,
the image magnifications of the Einstein cross type lenses are very weak
functions of source size up 
to $\sim 1\kpc$.  This is also true for the magnification ratios of
widely separated images in the fold and cusp caustic lenses.  This means
that selected magnification ratios for different the emission regions of a lensed
quasar should agree with each other, barring microlensing by stars.  The
source size dependence of the magnification ratio between the closest
pair of images is more sensitive to substructure.  
}


\newpage
\section{Introduction}
\label{introduction}
Evidence has been mounting  that there is a large amount of small-scale structure in the distribution of matter in the gravitational
lenses responsible for multiple images quasi stellar objects (QSOs)
\citep{Metcalf04,cirpass2237,KGP2002,2002ApJ...567L...5M,KD2003,Dalal2002,2002ApJ...565...17C,2001ApJ...563....9M,1998MNRAS.295..587M}.
The identity of this substructure is a matter of some debate.
 The Cold Dark Matter (CDM) model does predict that some substructure
 in the dark matter will survive within the halo of the lens galaxy.
 \cite{2004ApJ...604L...5M} have argued that, according to $\Lambda$CDM
 (CDM with a cosmological constant) simulations, there is not enough surviving
 substructure in the central regions of halos of galaxies to account for previous lensing
 estimates.  \cite{Metcalf04} has argued that, within the $\Lambda$CDM model, halos
 outside the lens galaxy will have enough of an effect to account for
 most of the observed lensing anomalies. This effect has also been investigated be \cite{2003ApJ...592...24C}.  It is important to determine just how
 important the substructure within the lensing galaxy will be.  In this
 paper we address this question by studying the properties of a
 gravitational lens taken  directly from a cosmological simulation.

Five important methods have emerged for detecting substructure in a
relatively model independent way.  Initially, the simple magnification
ratios were compared to predictions from a model for each lens
\citep{2002ApJ...565...17C,Dalal2002} or a family of models
\citep{2002ApJ...567L...5M}. 
This showed that most of the observed lenses are not consistent with
the simple lens models that are usually used and that
substructure was a probable explanation.  \cite{2001ApJ...563....9M}
\citep[also see][]{astro-ph/0109347,1995ApJ...447L.105W,2003ApJ...582...17K}
predicted that the presence of
substructure will cause changes in the image magnifications that are skewed
differently for negative parity (saddle-point) images than they are for
positive parity (minimum) images.  \cite{2002ApJ...580..685S}
investigated this in the case of microlensing and \cite{KD2003} 
showed that a sample of the observed QSO lenses do, in fact, show this
tendency even at radio wavelengths.  This is a property that is difficult
to reproduce by any possible contaminating effect and it is
relatively independent of the lens model used.

\cite{1998MNRAS.295..587M} pointed out that lens B1422+231 violated the
cusp caustic magnification relation \citep[see also][]{1986ApJ...310..568B,1990QJRAS..31..305B}.  If the source is
close enough to, and inside of, a cusp in a caustic curve, three of the
images will be clustered together. The sum of their magnifications
will be zero (taking the negative parity image to have negative
magnification).  \cite{KGP2002} showed that this relation holds for a
wide class of smooth, analytic lens models when three of the images
are close enough together.  They also showed that all known and well observed, cases
violate this relation although some of the cases might be due to
microlensing by stars rather than larger mass substructures.  Analyzing
simulated galaxies \cite{astro-ph/0306238} claimed that this relation
can be violated without substructure if the lens has a stellar disk.
In this paper we will revisit this question of how reliable the cusp
caustic relation is and what kind of substructure can cause violations.

Another method for avoiding model dependence was proposed by
\cite{MM02} and was put into practice by \cite{cirpass2237}.  In this
approach the QSO magnification ratios are measured in several different wave
bands.  Since the sizes of the emission regions of a QSO are strongly
dependent on wavelength (from $\simlt$1000\,AU for the visible to
hundreds of parsecs for the narrow emission lines) the magnification ratios could
potentially be different if there is structure in the lens (or
somewhere in front the lens or between the lens and the source) that
is of a similar size scale.
\cite{cirpass2237} found that, among other things,
the mid-infrared, radio and narrow line magnification ratios are not
consistent with each other and a lens without small-scale substructure.  A lower
limit was put on the mass and density of substructures that could cause
this mismatch.  The reliability of this method rests on the requirement that
the magnification ratios are not a strong function of source size.
\cite{cirpass2237} tested this assumption for a smooth analytic model
of the lens used in that study, but here we will test it with more
realistic lens models and in more generality (Section~\ref{sec:diff-magn-rati}).

The reliability of these methods for detecting
substructure in gravitational lenses has only been verified in
cases where the lens galaxy is represented by simple analytic models.
The one exception to this is \cite{astro-ph/0306238} where lenses were
simulated using the hydrodynamic, galaxy formation simulations 
\citep{2003ApJ...591..499A,2003ApJ...590..619M}. 
\cite{astro-ph/0306238}
investigated the cusp caustic magnification relation and the
statistics of even/odd parity image magnification ratios.  They found that the cusp
caustic relation could be grossly violated for disk
lens galaxies and that in the presence of substructures the caustics would
develop a large number of ``swallow tails''.  Their simulated galaxies
were not entirely realistic because of limitations in the simulations.
Most notably, the baryons and dark matter in their simulations are much
more concentrated in the center than in real galaxies.  This can
potentially have a large effect on the lensing properties because the
surface density at the location of the images is significantly
smaller than what it would be in a real lens. Furthermore, a concentrated
mass distribution makes the potential more spherical at the distance of
the Einstein ring.  Our results will be compared to this work.

The paper is organized as follows.  In section~\ref{sec:simulations}
the simulations are described in two parts starting with the galaxy
simulations and then the lensing calculations.
Section~\ref{sec:results} has the lensing results broken down in to
several subsections.  In section~\ref{sec:noiseSIE} we study the
noise in the image magnifications caused by shot noise in the N-body
simulations.  In section~\ref{sec:caustic-structure} we look at the
structure of the lensing caustics and the probabilities for different
image multiplicities.  The dependence of the magnification ratios on
source size is addressed in section~\ref{sec:diff-magn-rati} and in
section~\ref{sec:cusp-relation} the violations of the cusp caustic
relation for our simulated lens are studied.  A very brief comparison of
these results with observations is given in
section~\ref{sec:very-brief-comp}.  A summary of the paper and a
discussion of its wider implications are given in section~\ref{sec:concl--disc}. 

\section{Simulations}\label{sec:simulations}

To test the effects substructure has on galactic lenses, we adopt a hybrid 
approach, which implants an idealized model galaxy (either disk or elliptical) into
the center of a dark matter halo that has been extracted from a collisionless 
N-body simulation.  The advantage of this approach lies in the fact that we are 
not limited by the inability of current numerical simulations to generate realistic galaxies.  Conversely, the disadvantage is that the dark matter halo has been
evolved to redshift zero without the dynamical effects of a baryonic galaxy. 
Below, we describe how we attempt to overcome this limitation.

For all idealized model galaxies considered here, we extracted the
dark matter halo from the N-body simulation of
\cite{1999ApJ...524L..19M}.  This is a Milky Way sized galactic halo
and was simulated to a red-shift of z=0 with a force resolution of $\sim 0.5\kpc$  and contains 
1,362,104 dark matter particles within 265~kpc, each with a mass of 
$1.68\times10^6\msun$. This makes the total
mass within this box roughly $2.29 \times10^{12}\msun$.  The minimum resolved
substructure ($\sim$ 50 particles) is approximately
$8.4\times10^7\msun$, which is a factor of 6 better than that of
\cite{astro-ph/0306238}. 

\begin{figure}[t]
\begin{center}
\epsfxsize=15pc
\epsfbox{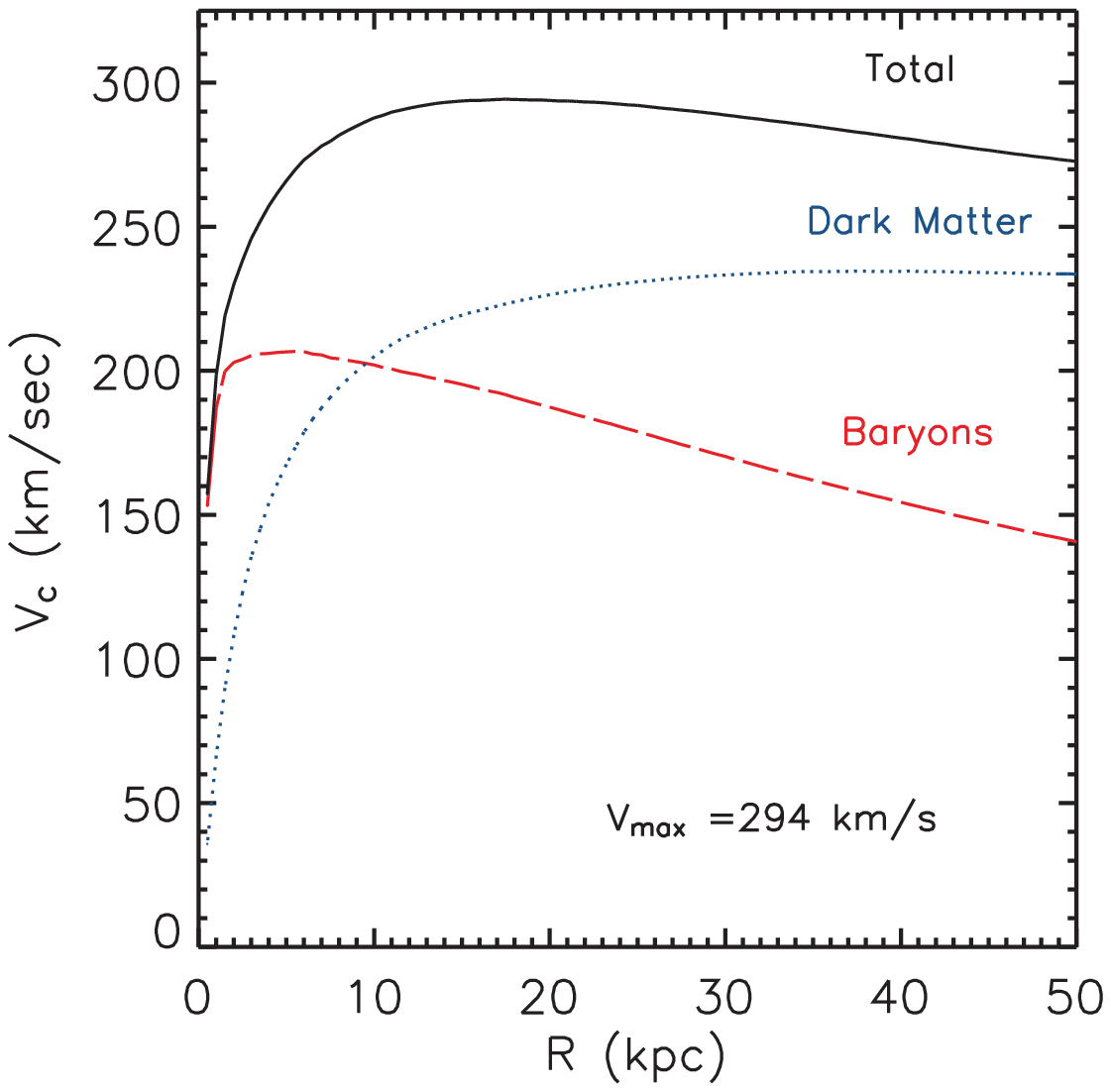}
\epsfxsize=15pc
\epsfbox{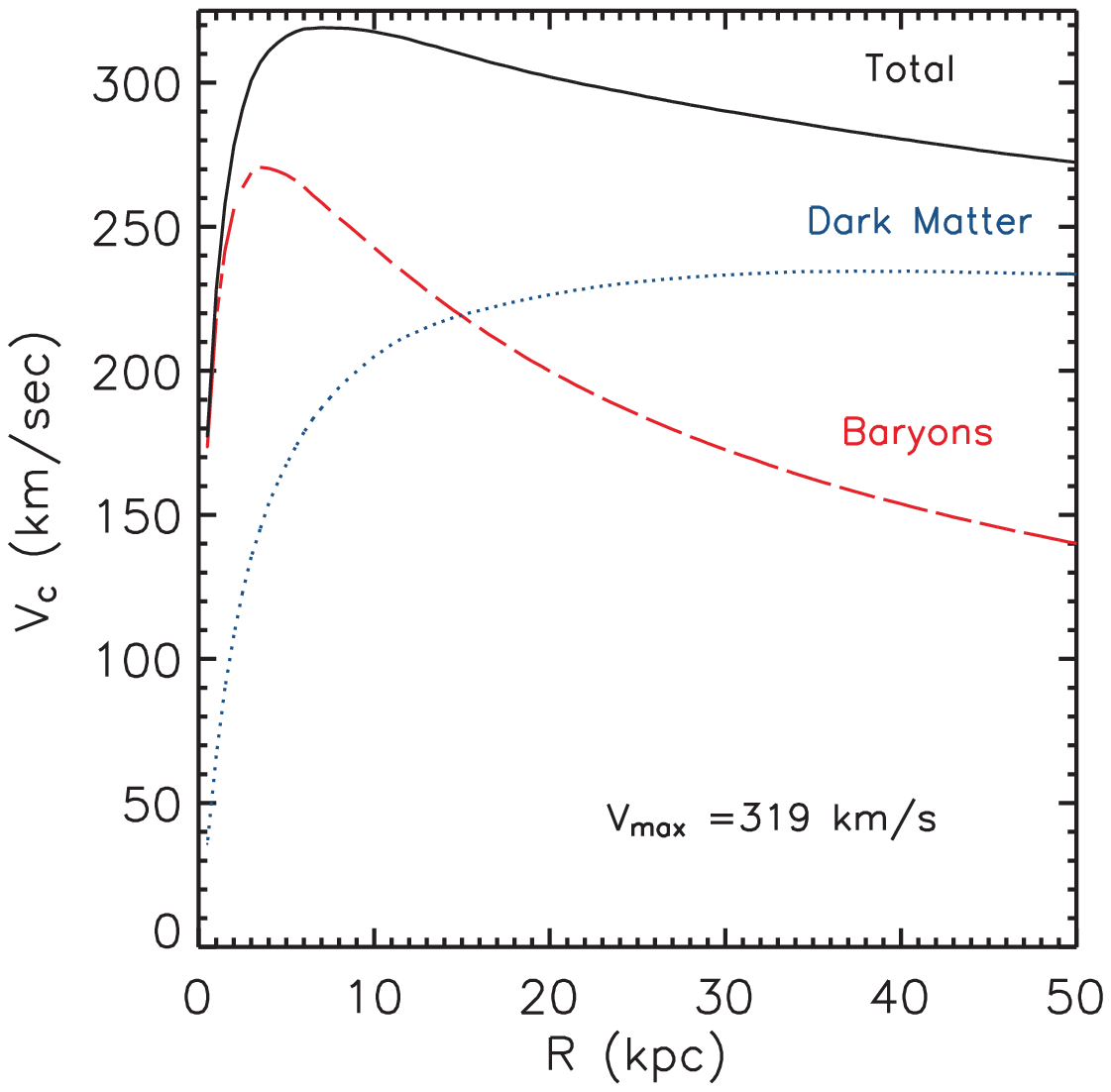}
\caption{\footnotesize   Circular velocities for our simulations with
  a disk galaxy implanted (left) and with an elliptical galaxy
  implanted (right) in the dark matter halo.  For each configuration
  contributions to the circular velocity from the dark matter and the
  baryonic matter are shown.}
\label{fig:velocity}
\end{center}
\end{figure}

As mentioned above, we implant two idealized model galaxies into the center of
our dark matter halo.  To conserve the total mass, we decreased
the mass of all dark matter particles in accordance with the implanted
baryonic mass.  The first idealized galaxy is modeled after the early-type
spiral lens galaxy Q2237+0305 \citep{2002MNRAS.334..621T}.  The total baryonic
mass implanted is $2.5 \times10^{11}\msun$, of which 71\% is in an
exponential disk with a scale radius of 9.5~kpc, and the remaining 29\% is in
the form of a \cite{1990ApJ...356..359H} bulge of effective radius
7.0 kpc.  The resulting circular velocity, figure~\ref{fig:velocity},
is very similar to figure 2 of \cite{2002MNRAS.334..621T}. Care was
taken to align the dark halo's angular momentum with the azimuthal
axis of the disk.

The second galactic system implanted is modeled after a generic
elliptical galaxy. To help in interpreting our results, we set the
total baryonic mass to be the same as in the early-type spiral system.
The elliptical is given a spherical Hernquist profile with effective
(half-mass) radius of 7~kpc and is subsequently deformed into a
tri-axial shape with projected ellipticities of 0.03, 0.16 and 0.33,
depending on the viewing angle. It is aligned with the tri-axial shape
of the unperturbed halo.  The rotation curve of this galaxy is also
shown in figure~\ref{fig:velocity}.

Each of our two implanted galaxies was realized with 50,000 particles and the
total system, baryonic galaxy plus dark matter halo, was run forward in time
using the N-body code GADGET \citep{2001NewA....6...79S} for 200 Myr.  
The baryonic particles were held fixed in position while the dark
matter halo was  
allowed to dynamically respond to the imposed galaxy potential.  Due to the
instantaneously imposed baryonic potential, the dark matter halo becomes more
spherical and contracts, increasing its central density.  Prior work
has suggested that our instantaneous application of the baryonic potential is
not much different then slowly growing it within a dark matter halo \citep{2002ApJ...571L..89J}. 

In contrast, the \cite{2002NewA....7..155S} simulations used by
\cite{astro-ph/0306238} in their lensing simulations are much more
concentrated than ours and than realistic galaxies.  For example the
disk component of their galaxies are only 3~kpc in diameter and when
compared to a real galaxy of similar rotation speed the angular
momentum of the stars is an order of magnitude too small
\citep[see][]{2003ApJ...591..499A}.

\subsection{Ray-Tracing}
\label{ray-tracing}

To study the lensing properties of our simulated galaxies, we must
determine the deflection angle as a function of position on the sky.
It is an excellent approximation in this case to treat the mass as if
it were in a single plane -- the lens plane.  A viewing angle is chosen
and the mass distribution is projected to find
the surface density, $\Sigma(\vec{x})$.
We can then calculate the dimensionless surface density or convergence
\begin{equation}\label{eq:kappa}
\kappa(\vec{x})=\Sigma(\vec{x})/\Sigma_{cr}.
\end{equation}
The critical surface density is
\begin{equation}\label{eq:crit}
\Sigma_{cr}=\frac{c^{2}D_{s}}{4\pi G D_{d} D_{ds}},
\end{equation}
where $D_{s}$ is the angular diameter distance between the observer and
the source; $D_{d}$ is the angular distance between the observer and the
lensing plane; and $D_{ds}$ the angular distance from lensing plane
 to source plane. We have chosen to put the source at z=3.0 and the
lens plane at z=0.8, which for a standard $\Lambda$CDM concordance
model ($\Omega_\Lambda=0.7$, $\Omega_m=0.3$)
with  $H_o=70\kms\mbox{Mpc}^{-1}$ gives
$\Sigma_c=1.92\times10^9 \rm \msun/kpc^2$.  Although the simulated
dark matter halo we use was simulated to $z=0$ we do not think that
repositioning it at $z=0.8$ biases our results in any significant way.
The variation in the substructure content in different halos is large
compared to the difference between the average substructure content of
halos at redshifts 0 and 0.8.  

The surface density can be related to the deflection potential,
$\psi(\vec{x})$,  through the Poisson Equation 
\begin{equation}\label{poisson}
\nabla^2\psi(\vec{x})=2\kappa(\vec{x}),
\end{equation}
where all derivatives are with respect to distance on the lens plane.
The potential, in turn, is related to the deflection angle $\vec{\alpha}$ by
\begin{equation}\label{alpha}
\vec{\alpha}(\vec{x})=\nabla\psi(\vec{x}),
\end{equation}
so that the position of the source on the lens plane, $\vec{y}$, is given by the
lens equation: $\vec{y}=\vec{x}-\vec{\alpha}(\vec{x})$.  The
magnification matrix is $A_{ij}\equiv \frac{\partial y^i}{\partial
  x^j}$, and the magnification of an infinitesimal image at $\vec{x}$ is
$\mu(\vec{x}) = |A|^{-1}$.  The two components of shear are defined as
$\gamma_1=(A_{11}-A_{22})/2$ and $\gamma_2=A_{12}=A_{21}$. 

We solve the above equations in discrete Fourier space. This process
involves first placing the discrete particle from the N-body
simulations on a regular two-dimensional grid, which was done using a
Cloud in Cell (CIC) routine (http://idlastro.gsfc.nasa.gov).  This
routine distributes the mass of each particle over the four pixels
closest to the particle position. The fraction of the mass in each
pixel is weighted by how close the particle lies to the center of the
pixel.  We used a  4096$\times$4096 grid at this stage.  After performing a
  Fourier transform on the discrete surface density we 
  calculated the lensing potential through the Fourier equivalent of
  equation~(\ref{poisson}).  The first and second derivatives of the
  lensing potential are also efficiently calculated in Fourier-space.  
  After transforming these quantities back to real-space the lensing
  properties (deflection angels, shear, magnifications, etc.) of our
  galaxies on the discrete grid. 

Solving the equations in Fourier space imposes unphysical periodic
boundary conditions.  To overcome this problem, we pad our simulations
with zeros on the borders.  By doing this we effectively separate our
galaxies from their mirror images to the extent that they do not
affect each others' strong lensing properties.  
The amount
  padding was chosen to overcome the problems introduced by two
  competing factors.  Reducing the effect of repeating boundary
  conditions favors a large amount of padding.  However, since we are
  limited by the overall resolution of  our ray-tracing, increasing
  the padding reduces the resolution at which we are able to sample
  our simulated galaxies. This in turn increases the errors introduced
  through pixelization. Using simple SIS and SIE models we find
that filling an  
array of 3052$\times$3052 with our simulated galaxies and filling in the 
rest of the 4096$\times$4096 grid with zeros gives the best
  compromise between the two competing factors.    
  We call this $\frac{1}{4}$ padding. The case where 2048 the grid is filled with the simulation and the rest with zeros we call $\frac{1}{2}$ padding.  If the padding is
  increased from $\frac{1}{4}$ to the $\frac{1}{2}$ level, then the errors in the deflection angle near the critical curve ($\kappa =0.5$ for the SIS and SIE) increases from roughly 0.8\% to 2\%.  This is because in this inner region, the $\frac{1}{2}$ padding becomes dominated by the pixel scale.  We also find that reducing the padding below $\frac{1}{4}$ does not significantly improve accuracy at the critical curve but instead leads to an increase in errors as we move out away from this region.  In the $\frac{1}{4}$  case, the errors stay at the same levels throughout the strong lensing region.  Similar tests were also performed for the magnification.  We found that the magnification is even more sensitive to the pixel scale.  For regions within the critical curve, errors in the magnification were found to be below 5\% for the $\frac{1}{4}$ case but at the 20\% level for $\frac{1}{2}$ padding.

The deflection angle, Equation~(\ref{alpha}), is calculated at uniform grid
points on the lens plane and the lens equation allows us to map each
image pixel back to a point on the source plane.  Nonetheless, mapping
back from the source plane to the image plane is not simple, because
each source position can be mapped to many image positions. To overcome
this difficulty, a number of techniques have been developed. At first we pixelize the source plane and associate image plane
pixels with the source plane pixels closest to their position when
mapped back to the source plane. Through this method, we are able to
generate a list of all image pixels associated with any source plane pixel.

The 4096$\times$4096 grid does not provide enough pixels in the inner regions 
affected by strong lensing to model lensing properties accurately. Once we have calculated lensing properties such 
as angular deflections, shear and magnification on the grid, the resolution is 
increased by extracting the central region of interest in strong lensing 
and using bilinear interpolation to calculate these quantities to higher 
resolution.  This is justified because the numerical resolution of
the simulations is already 30\% larger than the
4096$\times$4096 grid resolution. Therefore any structure on a smaller scale
would be purely numerical noise in any case.

The critical curves on the image plane are where the magnification diverges 
to infinity.  Close images will always be separated by one of these
curves.  Images on opposite sides of a critical curve have 
different parities -- their magnifications, defined as
$1/\det[A]$,  have opposing signs.  We
calculate the magnification at each image pixel and use this fact to 
separate images that are close to critical curves. 
In addition, mapping these critical lines back to the source plane
with the lens equation allows us to map out the caustic curves, which are the
curves on the source plane where the magnification diverges.  The caustic
curves also separate regions on the source plane with different
image multiplicities. 

Using the sign of $\det[A(x)]$ to separate images works well for much of
our work, but we found that for small sources near a caustic this
method requires 
very high grid resolution and becomes too computer intensive.
For these circumstances, we developed an adaptive mesh technique, which allowed
us to study source sizes well below the resolution of the grid on
which the lensing equations are solved. 
After extracting the inner regions of the lens we then use the
  deflection angle information to map a coarse grid to an irregular
  grid on the source plane.  By selecting a source position, we
  calculate its distance from each of the grid points on the source
  plane.  This forms a 2D scalar field on the image plane.  We then
  are able to identify the positions of the images 
formed by this source by finding the minima of this field.  We then
extract regions around each of the image 
positions and use interpolation techniques only on these small
regions, thereby greatly reducing the computational demands of the
calculation.  This refinement procedure can be repeated multiple times
to obtain higher resolution when needed.
This allows us to cover the three orders of
magnitude in length scale required for some of the results that discuss
later.

Once the images had been identified, we investigated three methods for
measuring the magnifications of the images.  The most straightforward
method is to take the ratio of image size to the source size by counting
the number of pixels in each image.  A second method is to take the
average of the inverse of the magnification calculated at each pixel on
the image plane that is in the image.  The inverse of this is an
estimate of the magnification.  In the third method, the image
of the boundary of the source is found by interpolation.  The area on
the image plane enclosed in this curve is then calculated by integrating
$\vec{r}\times d\vec{r}/d\theta$ around the curve.  We find that for
isolated images, where none of the images come in contact with the
critical curves, all methods agree well with each other. However, as the
images grow and start to come into contact with the critical curves, the first method
becomes less accurate due to the limited grid resolution and the
extreme elongation of the images.  We also found that, in extreme cases, the
results of the boundary method can be dependent on the interpolation
algorithm used.  For these reasons, we chose to use the averaging method
in all our reported calculations.

\section{Results}
\label{sec:results}
\subsection{Noise Properties with a Singular Isothermal Ellipsoid (SIE)}
\label{sec:noiseSIE}

\begin{figure}
\begin{center}
\epsfxsize=30pc
\epsfbox{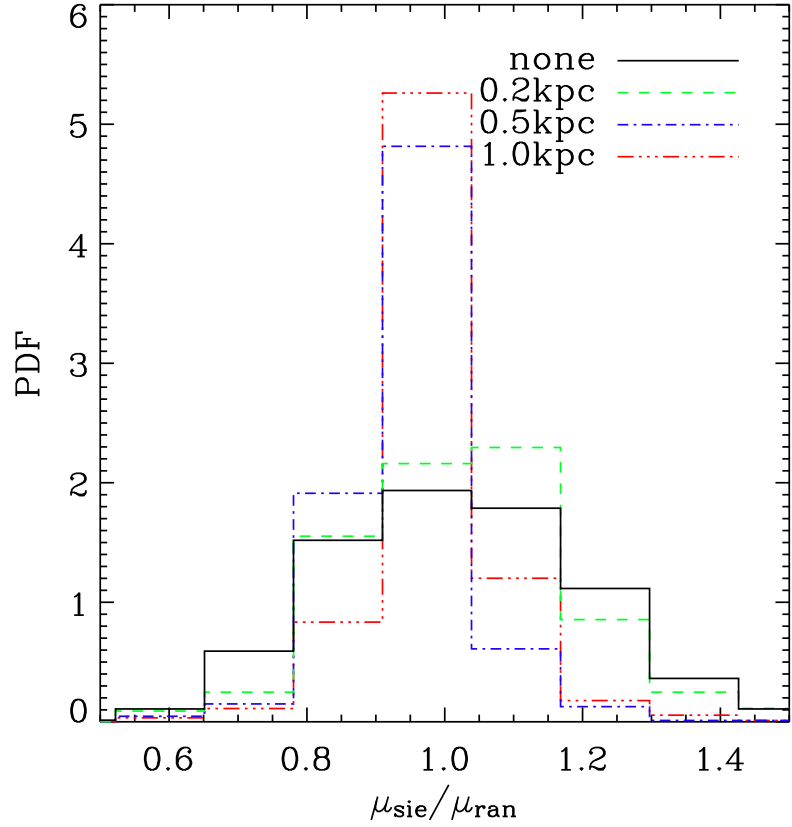}
\caption{\footnotesize  The Probably Distribution Function (PDF) for
  the ratio of the magnifications of images formed in the Singular Isothermal Ellipsoid relative to
  the corresponding images in the realization with random noise.  We
  see that without smoothing (solid black line), the random shot noise introduces a large
  spread in magnification.  The histograms are labeled with the width
  ($2\sigma$) of the Gaussian used to smooth the surface density.
  We see that smoothing with a width of 0.2~kpc or below does not
  have a significant effect.  To reduce the effect of noise to a
  sufficient level we need to use a smoothing scale of 0.5~kpc.  The standard deviations of the PDF with no smoothing, 0.2~kpc, 0.5~kpc and 1.0~kpc are $\sigma=$0.23, 0.21, 0.12 and
  0.12.}
 \label{fig:sie_pdf1}
\end{center}
\end{figure}

Because of the finite size of the particles in the simulations there
is a significant amount of shot noise in the surface density estimate
that in turn affects, the lensing properties.  We will remove this 
noise by smoothing the surface density with a Gaussian kernel, but
first we need to calibrate the process by applying it to a case where
the lensing properties are already well known.  To do this, we add
noise to a simple SIE lens model and repeatedly smooth it to find the
minimum smoothing length that reproduces the expected image
magnifications with sufficient accuracy.

First, we calculate the surface density of a SIE with caustic features
similar to those seen in our simulated halo (figure~\ref{fig:inc}).   
We then create different realizations of the surface density with the level of shot
noise we would expect if the SIE were made up of the same number of
particles as our simulated halo.  
We do this by adding noise directly to the 2D mass distribution.  The shot noise is reproduced by
first creating an array of normally distributed random numbers and
then convolving it with a two dimensional kernel designed so that the
second order correlations between adjacent pixels are the same as those
expected in the CIC estimate of the surface density with Poisson
fluctuations in the particle numbers.  Since the particle
density is high in all the regions of interest, we expect that
the central limit theorem holds and that the noise will be close to Gaussian.
There is no guarantee that the Poisson assumption will give an accurate
estimate for the magnitude of the shot noise.  As will be commented on
later, when a random realization of this 
noise is added to a real simulation the surface density tends to get
less smooth and the lensing properties more irregular.  For this reason
we suspect that this is an overestimate of the noise.

We first look at the affects on critical curves and caustic structure.
A true SIE lens does not have a radial (inner) critical curve and the
associated curve on the source plane is called a cut instead of a caustic.  This
is the result of the singularity in the density distribution at the
center of the lens.  Our simulated SIE will necessarily have a radial critical
curve and caustic because of limited resolution.  In effect the
simulated lens has a core which shrinks the size of the radial
caustic.  We find that the radius of the radial caustic is
underestimated by $\sim 37\%$ relative to the analyticly determined
cut.  This calls into question how well the simulations can predict
the existence and the magnification of any image that is close to the
center of the lens.  For this reason we do not believe that we can
accurately predict the relative probability of two image lenses to
four image lenses.  Such a calculation would also need to include the
magnification bias which strongly suppresses the likelihood of
observing an image very near the center of the lens.  How the
numerical uncertainties and magnification bias affect the image
multiplicities will need to be studied in more detail.
In contrast, the tangential caustic and critical curve are very well reproduced in
the SIE simulations.

We find that the addition of shot noise does not significantly alter
the shape of the caustics, but it does seem to affect the outer (tangential)
critical curve.  We see the same thing in
 section~\ref{sec:caustic-structure} for the simulated lens.  This
 outer critical curve is  
close to the $\kappa=0.5$ contour which is a region of much lower 
density than the inner (radial) critical curve.   This may explain the 
outer critical curve's higher sensitivity to shot noise, which can
also be seen in figure \ref{fig:smooth1}.  Since the magnification
becomes very large near the caustic curves, the variations in 
outer critical curve could have significant effects on image
magnifications (further examined in sections~\ref{sec:diff-magn-rati} 
and \ref{sec:cusp-relation}).  We investigate
this by choosing one hundred random source positions that lie within
the caustics.  For each of these positions we calculate 
the positions and magnifications of the images formed in each
realization and compare the magnifications to the SIE model.  
Significant discrepancies are seen before smoothing.  We smooth each
of the three SIE realizations with a Gaussian kernel
of widths (2$\sigma$) of 0.2, 0.5 and 1.0~kpc giving a total of 12
realizations (including the original SIE without smoothing).  For
each smoothing scale, we study the variation in magnification of the
images formed from the one hundred source positions.
The results are shown in figure~\ref{fig:sie_pdf1}.    With a smoothing
scale of 0.5~kpc, the magnification has a standard deviation ($\sigma$)
of 0.12, which we judged to be small enough 
for our investigations and we adopt it in later sections.
Smoothing on this scale does not significantly alter the sizes of
  the caustic regions as shown in figure~\ref{fig:smooth1}. Although
  it is clear that over smoothing will affect the overall profile of
  the lens, we believe that the level of smoothing we use does not
  have a large effect of our galaxy profile.  This is also
the smoothing scale for the N-Body code used to construct the dark
matter halo.   Therefore,  we can use this level of smoothing without losing any
real spatial information.  It should be noted, however, that the same
amount of smoothing will not result in the same variance in the
magnification ratios for every source location.  For some source
locations the magnifications could be more or less sensitive to shot noise.

\subsection{Caustic Structure}
\label{sec:caustic-structure}

\begin{figure}
\begin{center}
\epsfxsize=30pc
\epsfbox{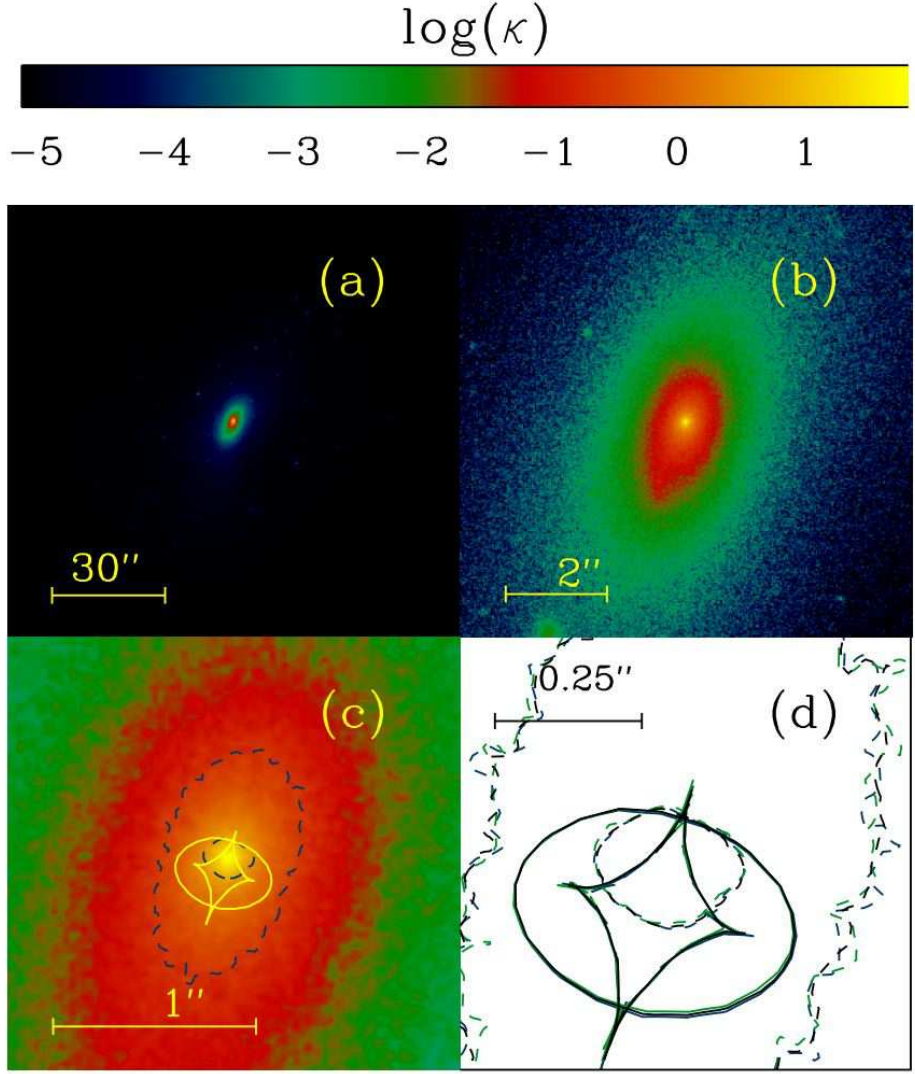}
\caption{\footnotesize  The simulated galaxy with the disk inclined at
  a 30 degree angle to the line of sight.  Figure~(a) shows the
  surface density of the entire
  galaxy along with the buffering we use to make sure the repeating
  boundary conditions do not affect our results. In (b) we see the
  inner region that is of interest to us.  Here there is some substructure
  which affects some of the overall lensing properties, such as the
  offset of caustics to density peaks as seen in (c).  In (c) the
  critical curves are shown as dashed curves and the caustic curves
  are shown in yellow.  The critical curves clearly have small-scale
  irregularities.  In (d) we look at the inner caustics region and how it is
  affected by random noise.  Results for three realizations of the
  shot-noise are shown in black, green and blue.
  Without smoothing, shot-noise does affect
  the critical curves, which, in turn are mapped back to the caustics.  We
  see some evidence for swallow tail features in the caustics, but they are not
  prominent and the inner (radial) caustic is relatively stable. No
  smoothing of the surface density has been done here. In the inner
  regions, where multiple images form, roughly 2/3 of the surface
  density is due to dark matter where as the remaining 1/3 is due to
  baryons.}
\label{fig:inc}
\end{center}
\end{figure}

\begin{figure}
\begin{center}
\epsfxsize=30pc
\epsfbox{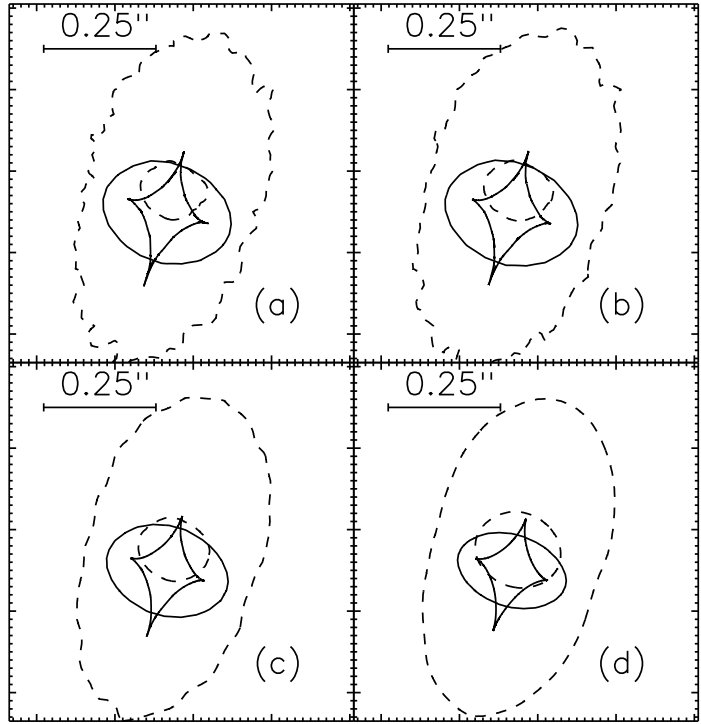}
\caption{\footnotesize   The effects of smoothing on caustic
  critical curve structure.  Panel (a) shows the results with no smoothing. Panels
  (b), (c) and (d) show the caustic structure and critical curves with
  Gaussian smoothing on scales 0.2, 0.5, 1.0~kpc, respectively. We see
  that the general shape of the outer critical curves (and hence the
  asteroid caustic which corresponds to it) is largely unaffected by the
  smoothing.   The smoothing, however, does affect small-scale
  variations of these lines, making them more smooth.  The radial (inner)
  critical curve (and the corresponding oval caustic curve) is affected in exactly
  the opposite way.  On small-scales, the noise and smoothing do not
  affect the curves.  This is because the regions close to the inner
  critical curves have a high density, hence they are less sensitive to
  shot noise.  The Gaussian smoothing, however, does cause the inner
  critical curve to expand and the oval caustic region to shrink.  This
  is because smoothing flattens the inner cusp region of the galaxy.}
\label{fig:smooth1}
\end{center}
\end{figure}

\begin{figure}
\begin{center}
\epsfxsize=30pc
\epsfbox{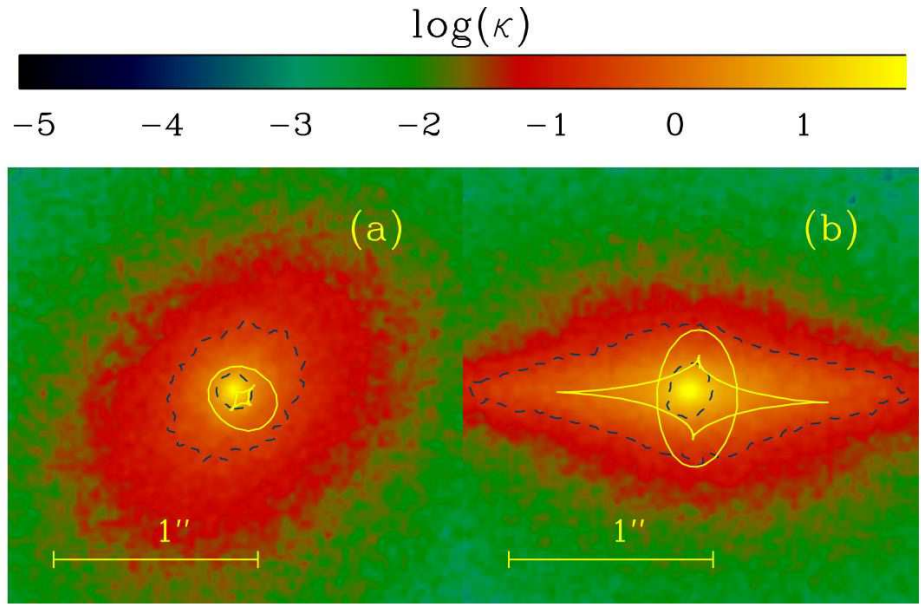}
\caption{\footnotesize On the left, (a),  shows the surface density,
  critical curves and caustics for the face on disk and on the right,
  (b), shows the same quantities for the edge on disk.  Between these
  two extreme cases we can see distinct changes in the caustic
  features.  In the face on case, where the surface density is almost
  circularly symmetric the inner, four image, caustic is small.  In
  the edge on case, one effect of the disk is to extend the inner
  caustic.  This inner caustic is so enlarged that we see large naked cusp
   regions where the inner caustic extends out beyond the outer caustic
   and three images are observable.  The outer caustic could be
   affected by limited resolution which makes it smaller than it would
   be with higher resolution.}
\label{fig:face_edge}
\end{center}
\end{figure}

\begin{figure}
\begin{center}
\epsfxsize=30pc
\epsfbox{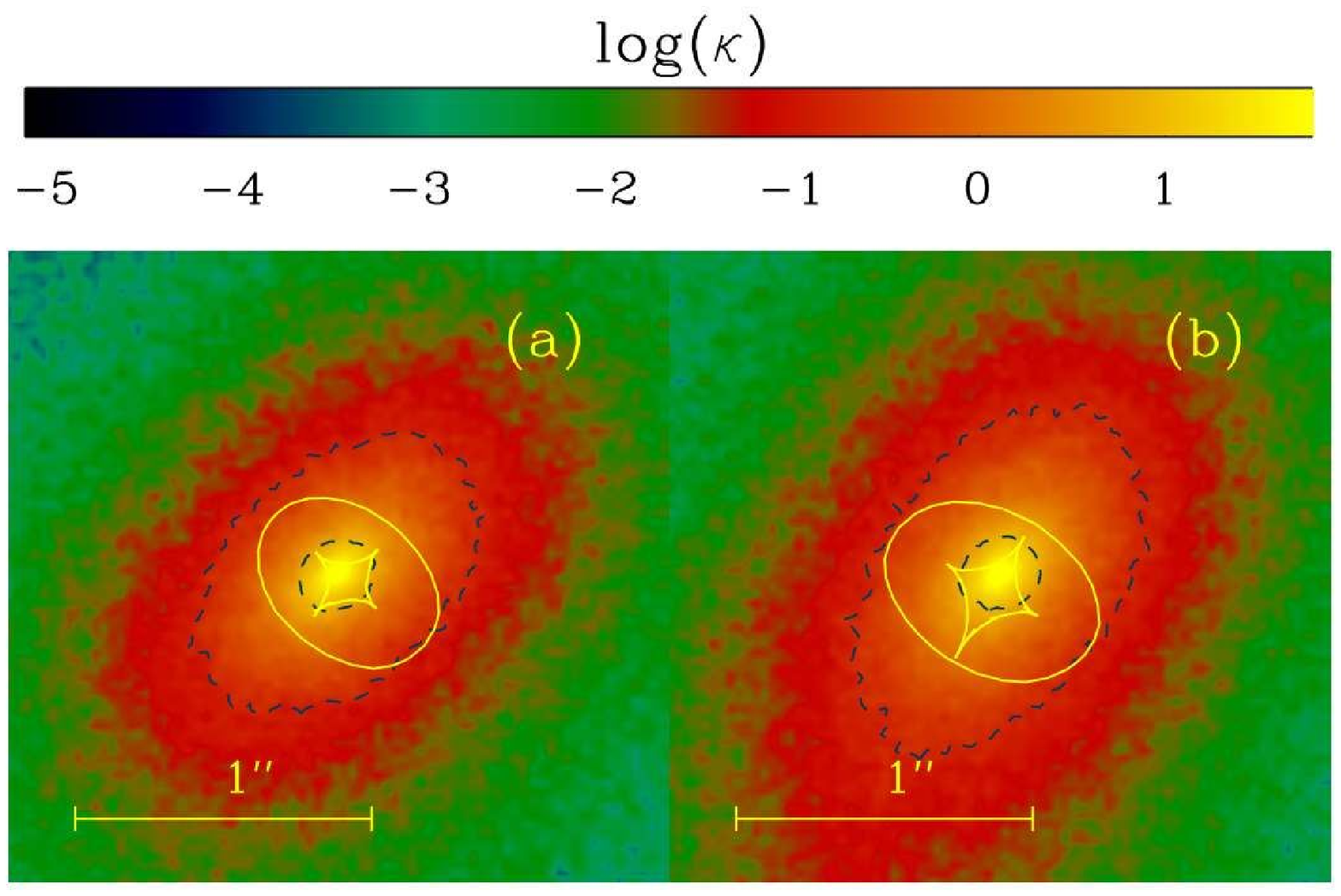}
\caption{\footnotesize The surface density, critical curves and
  caustics for two orientations of the elliptical galaxy - a)
  orientation~1 and b) orientation~2.
 We see that the elliptical galaxy does not show the large extended inner caustic
  that we see for the disk case.  This shows that the elliptical surface
  densities are more circularly symmetric, thus making it very difficult to
  create three image systems.} 
\label{fig:ellip}
\end{center}
\end{figure}

Here we show the caustic structures, critical curves and convergence maps 
for a number of viewing angles through our disk galaxy.  Figure~\ref{fig:inc} shows the case of our disk galaxy inclined at an angle
of 30 degrees to the line of sight.   
The result of smoothing the surface density is shown in figure
\ref{fig:smooth1}. 
We find that the noise does affect the outer critical curve, which
is the image of the inner (tangential) caustic curve.  
Noise has the greatest effect close to the cusps in the caustic, where we see
some 'swallow tail' structure as seen by \cite{astro-ph/0306238}.
Some of these features persist through the different realizations of
the noise indicating that they are real properties of the lens.
However, we do not see as many of these features as \cite{astro-ph/0306238} do.  This could
be because their simulated lens is much more concentrated and the density
at the critical curve is lower which increases the shot noise. In addition their particle mass is larger.

In figure~\ref{fig:face_edge}, we see the convergence, critical curves
and caustics for two cases: one with the disk face on to the line of
sight and the second with the disk edge on to the line of sight.  In
these images we see a large variation in the area of the two-image
regions and the four-image regions (not counting the central
demagnified image).  In the edge-on case, we also see 
the cusps of the tangential caustic extending out beyond the
radial caustic, forming what is called a {\it naked cusp}.  If the
source is in this region, three observable images and no central image
are formed.  As discussed in section~\ref{sec:noiseSIE} limitations in
resolution cause the area within the radial caustic to be
underestimated.  The radius of the inner critical curve is only about twice
the force resolution.  This prevents us from making an accurate
measurement of the relative numbers of four image to two image and
three image systems at this time.  In addition, any such predictions should include
the effects of magnification bias which could change the result by
more than a factor of 10
\citep{2002ApJ...577...51F,2004ApJ...608...25C,2001ApJ...553..709R,1996MNRAS.282...67K}.
While the radial caustic and critical curve do appear to be affected
by resolution limitation the tangential caustic, critical curve and
the magnifications of images near the tangential critical curve
(Einstein radius) are much more accurately calculated as discussed in section~\ref{sec:noiseSIE}.

In figure \ref{fig:inc}, we see that the caustics are not centered on
the density peak.  This is due to the lack of symmetry in the surface
density and in particular to the presence of a substructure just below
the density peak visible in figure \ref{fig:inc}, Panel (b).  We also
look at this offset in our 50 random realizations and find a mean
offset of 0.08 arc-seconds.

\subsection{Differential Magnification Ratios}
\label{sec:diff-magn-rati}

\begin{figure}
\begin{center}
\epsfxsize=12pc
\epsfbox{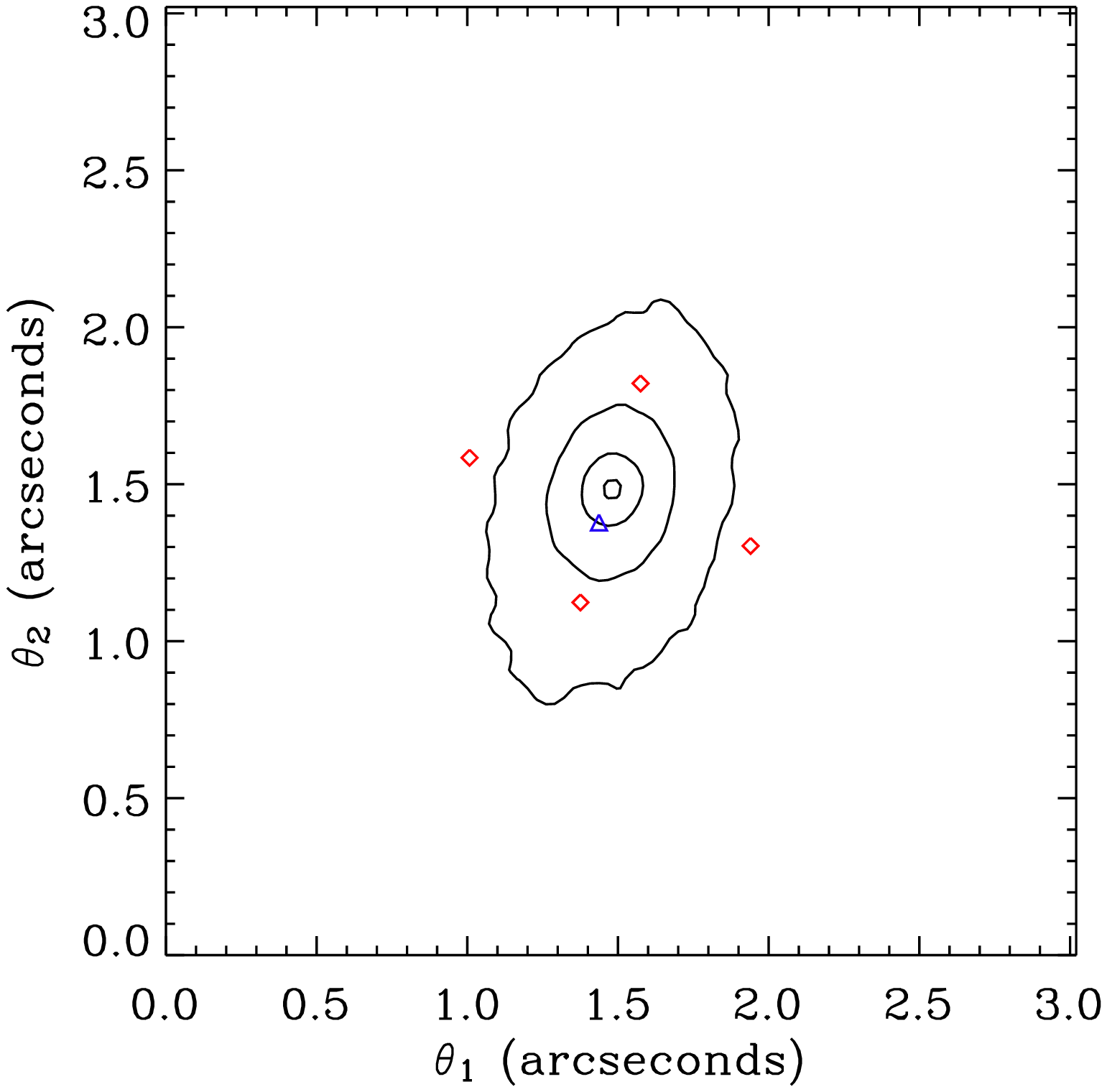}
\epsfxsize=12pc
\epsfbox{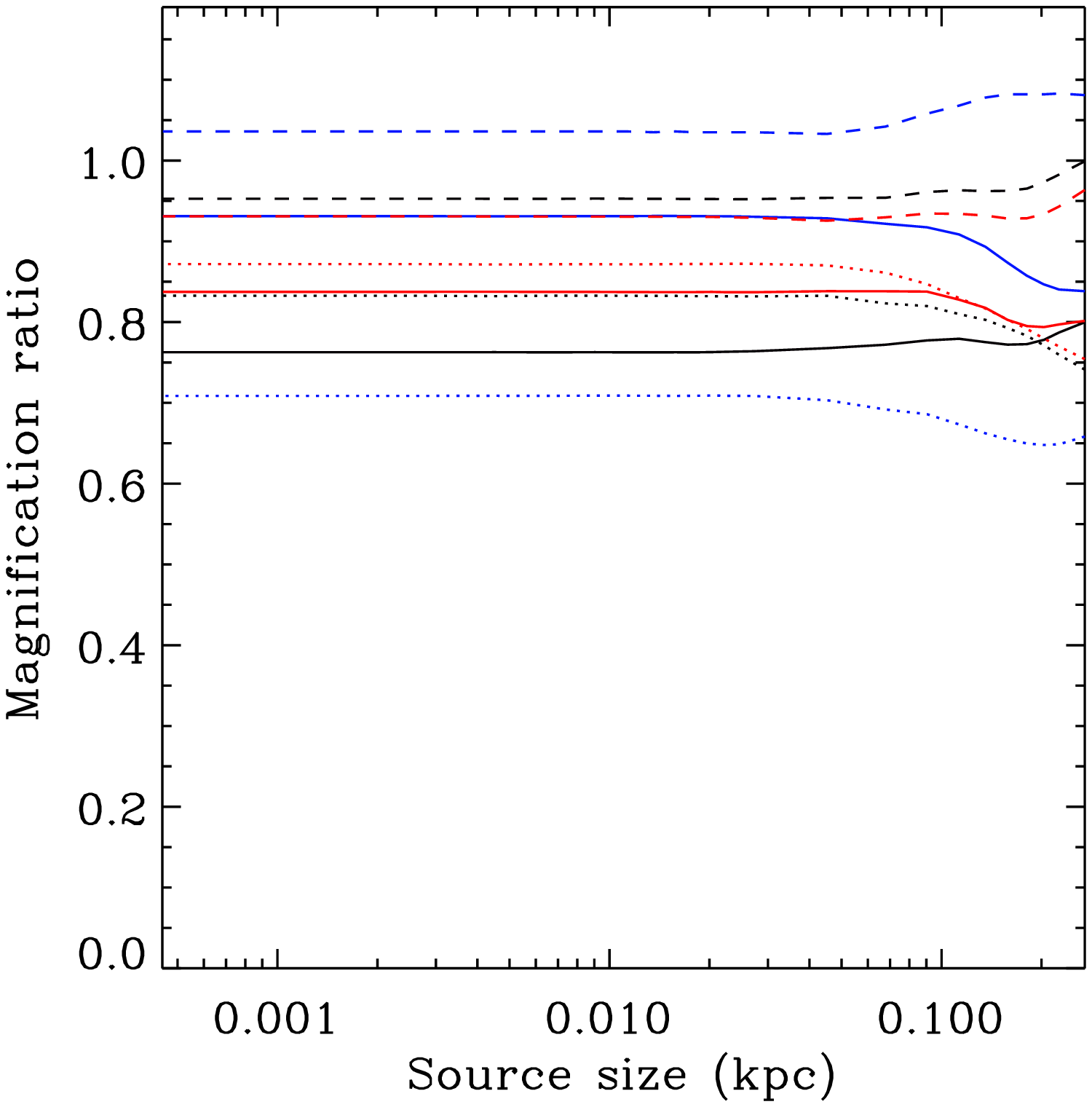}
\epsfxsize=12pc
\epsfbox{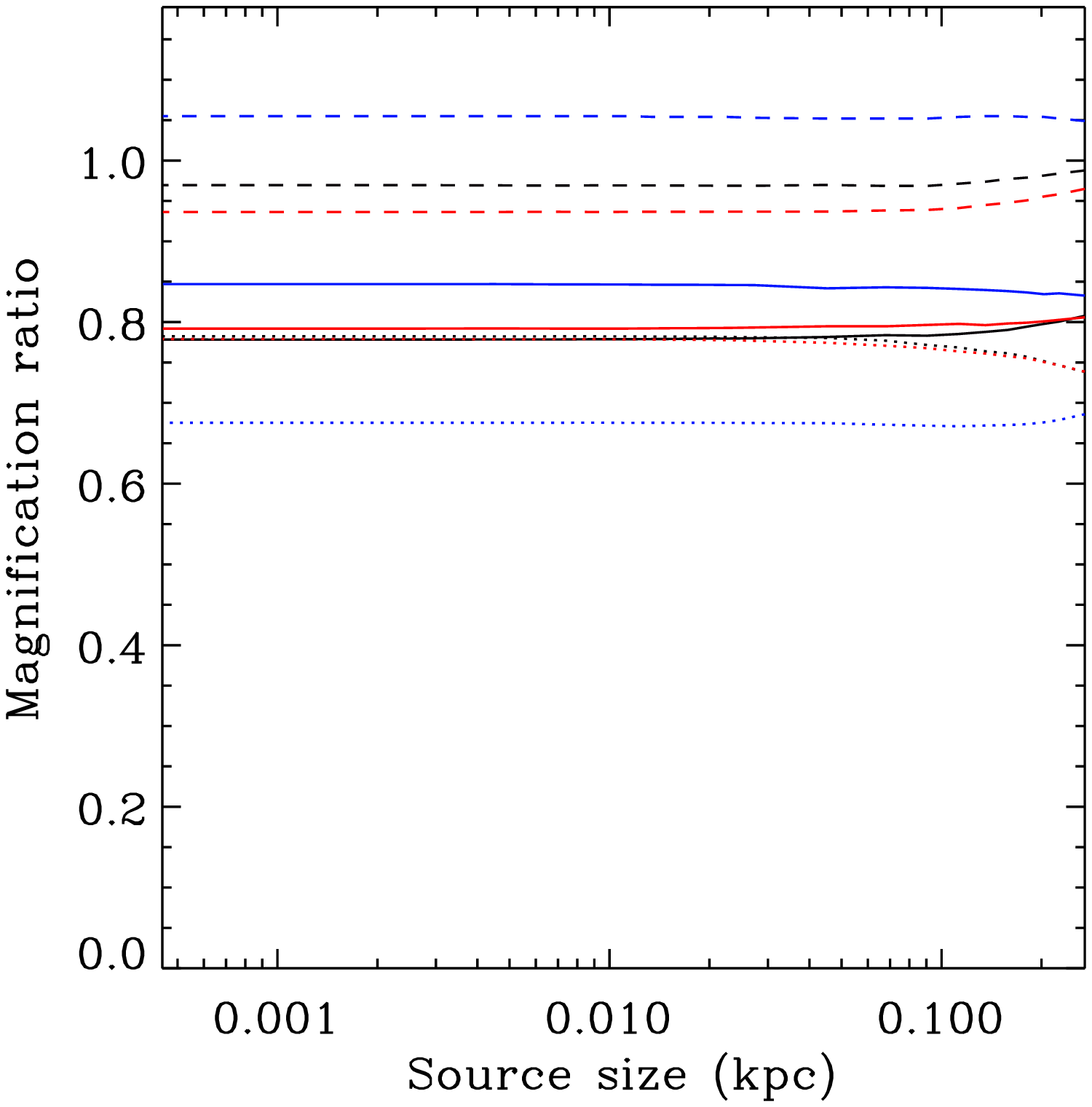}
\caption{\footnotesize On the left is the image configuration
  for an Einstein cross formed when the source is near the center of the
  radial caustic for the disk galaxy case inclined at 30$^\circ$.  The four bright images (red
  diamonds) correspond to the source position indicated by the blue
  triangle.  The contours show the surface density (with Gaussian
  smoothing = 0.5~kpc), the levels are $\kappa$ = 0.5,1.0, 2.0 and
  4.0.  The two panels on the right show magnification
  ratios as a function of source size.  The middle panel shows results {\it without
    smoothing} and the right panel shows the ratios when the surface density
  smoothed with a Gaussian kernel of width $2\sigma=0.5\kpc$. The
  solid curves show the magnification ratios of the closest image
  pair, the dotted curve shows the magnification ratio of the other
  two images, and the dashed curves shows the ratio of the average of
  each pair (see discussion in \S~\ref{sec:diff-magn-rati}).  The
  black lines are calculated using the n-body 
  simulations where as the red and blue lines are calculated after two
  realizations of the estimated noise has been added.}
\label{fig:magratio_cross}
\end{center}
\end{figure}

\begin{figure}
\begin{center}
\epsfxsize=15pc
\epsfbox{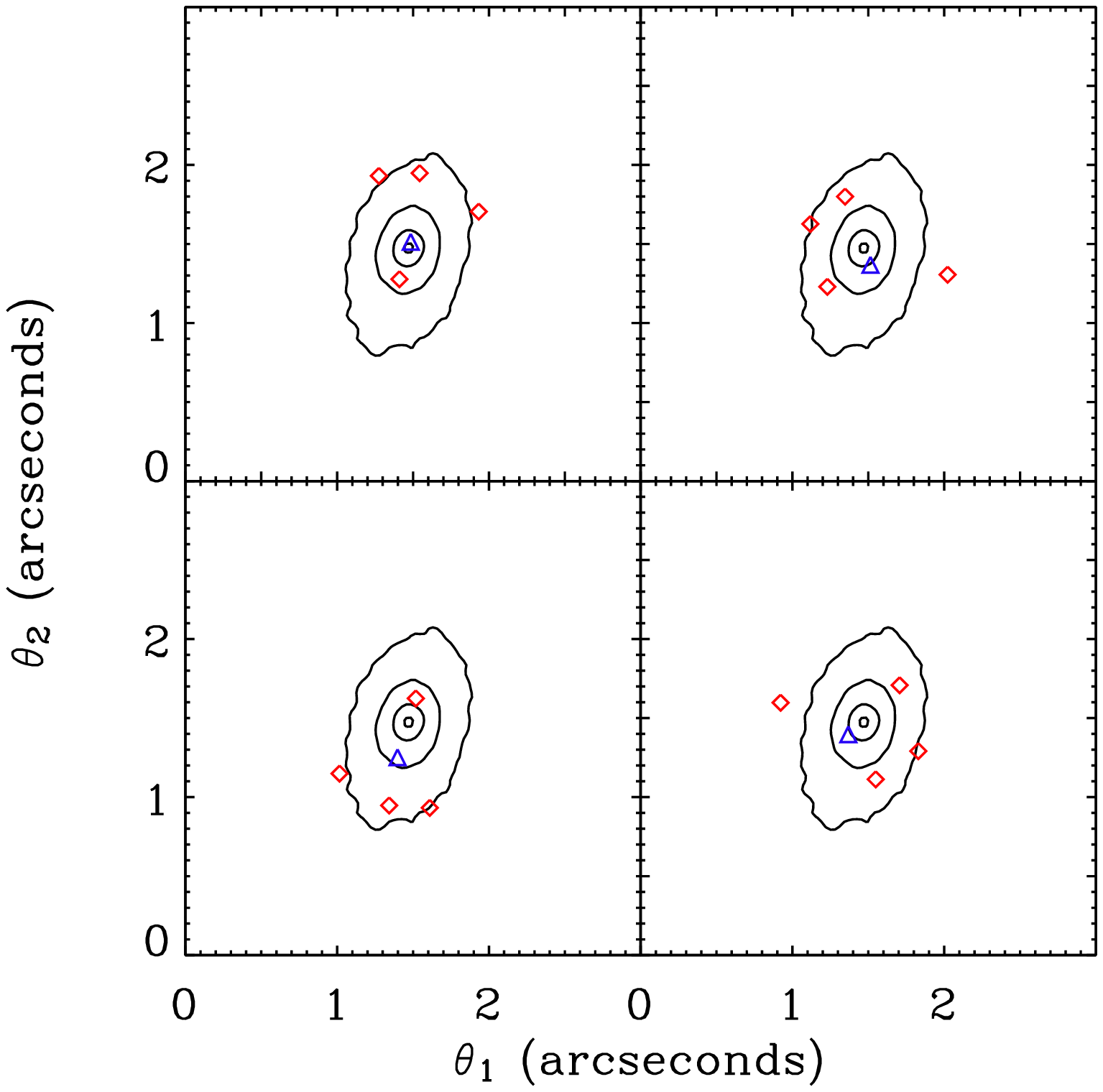}
\epsfxsize=15pc
\epsfbox{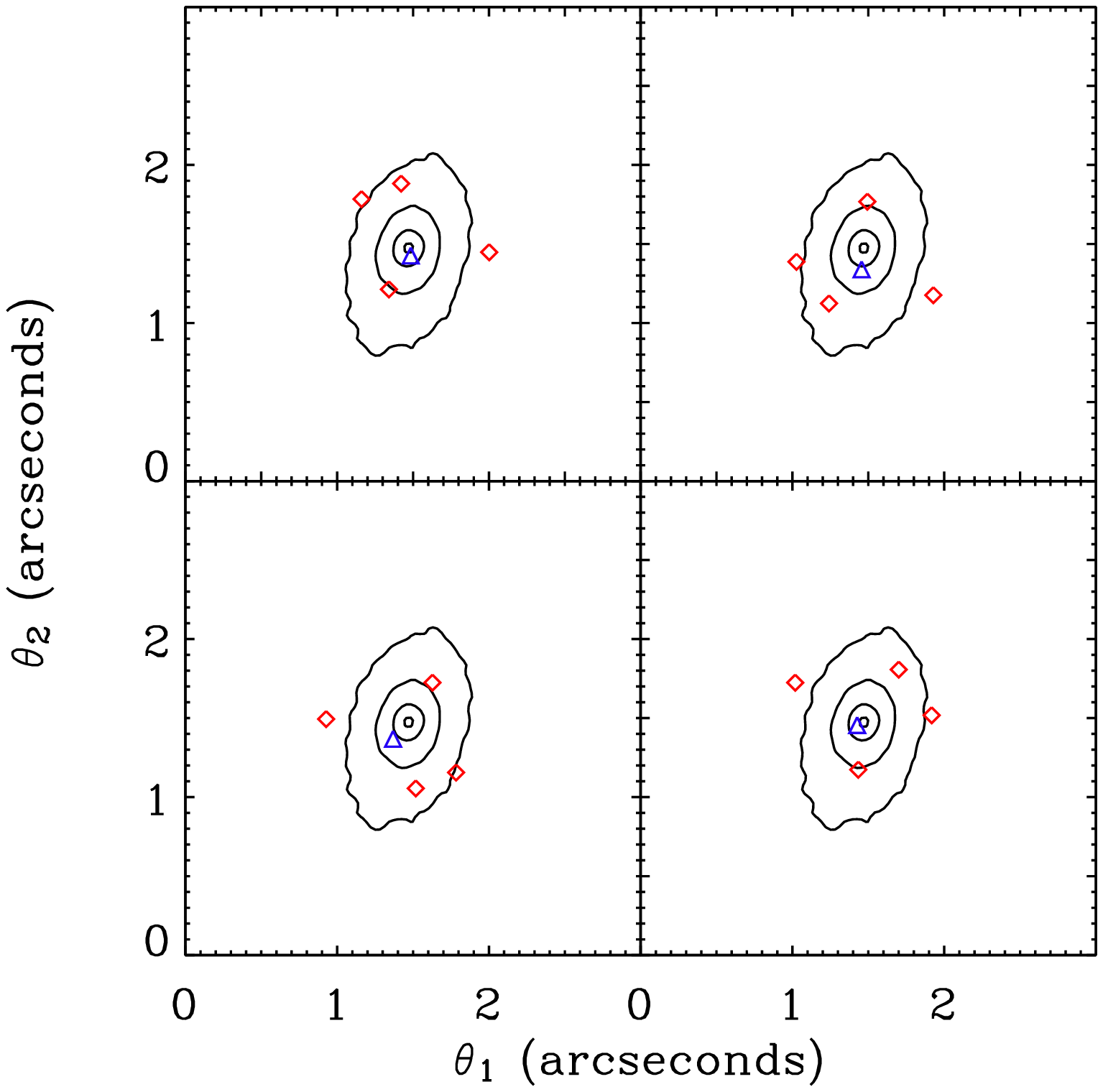}
\caption{\footnotesize   The image configurations used to test the
  dependence of the magnification ratios on source size.  On the left
  are four configurations with the source close  
  to a cusp in the caustic and on the right are four configurations with the
  source close to a fold caustic.  As in
  figure~\ref{fig:magratio_cross} red diamonds mark the images and
  blue triangles mark the source position.  The contours are as in
  figure~\ref{fig:magratio_cross}}
\label{fig:magratio_pos}
\end{center}
\end{figure}

\begin{figure}
\begin{center}
\epsfxsize=15pc
\epsfbox{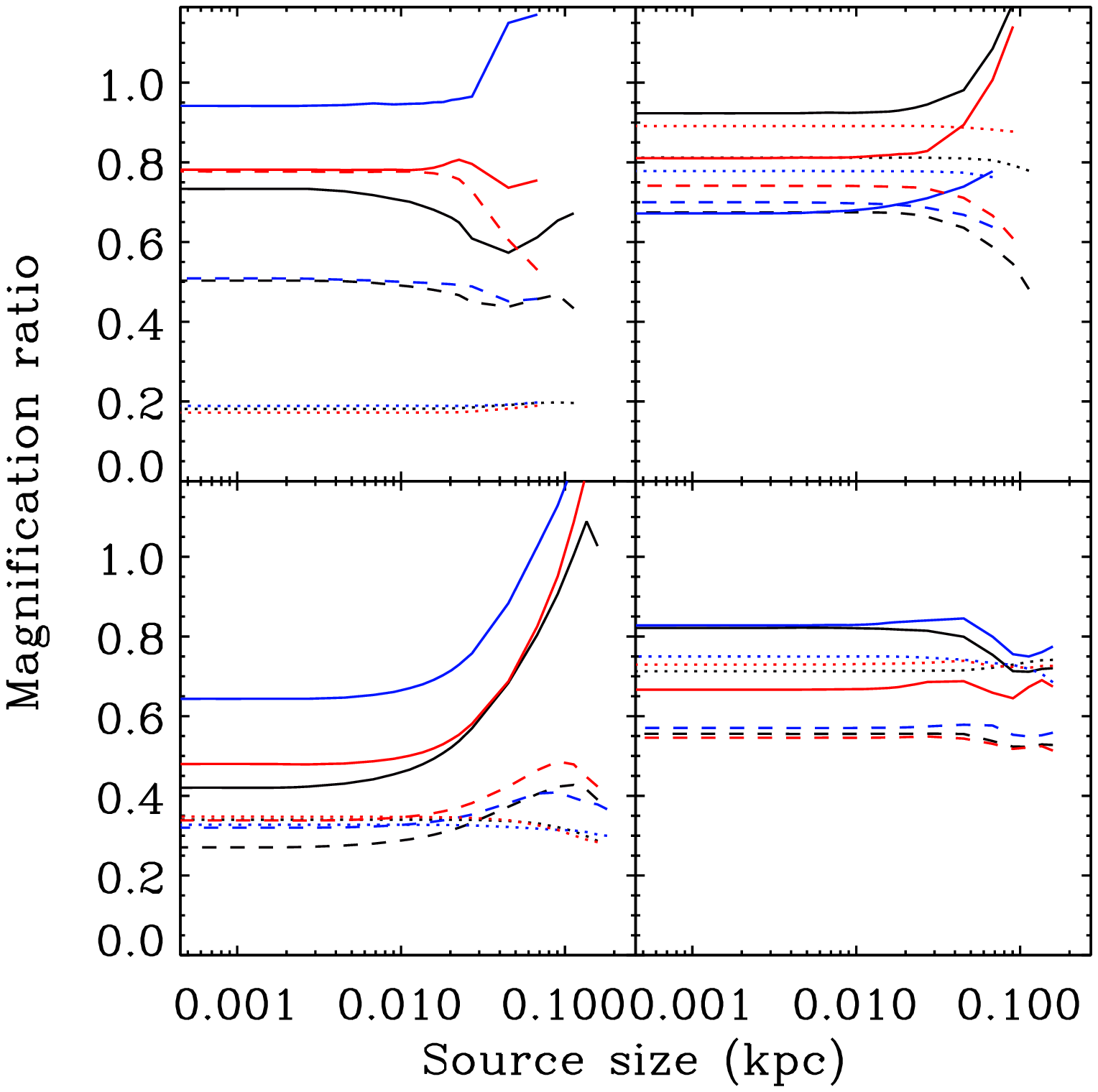}
\epsfxsize=15pc
\epsfbox{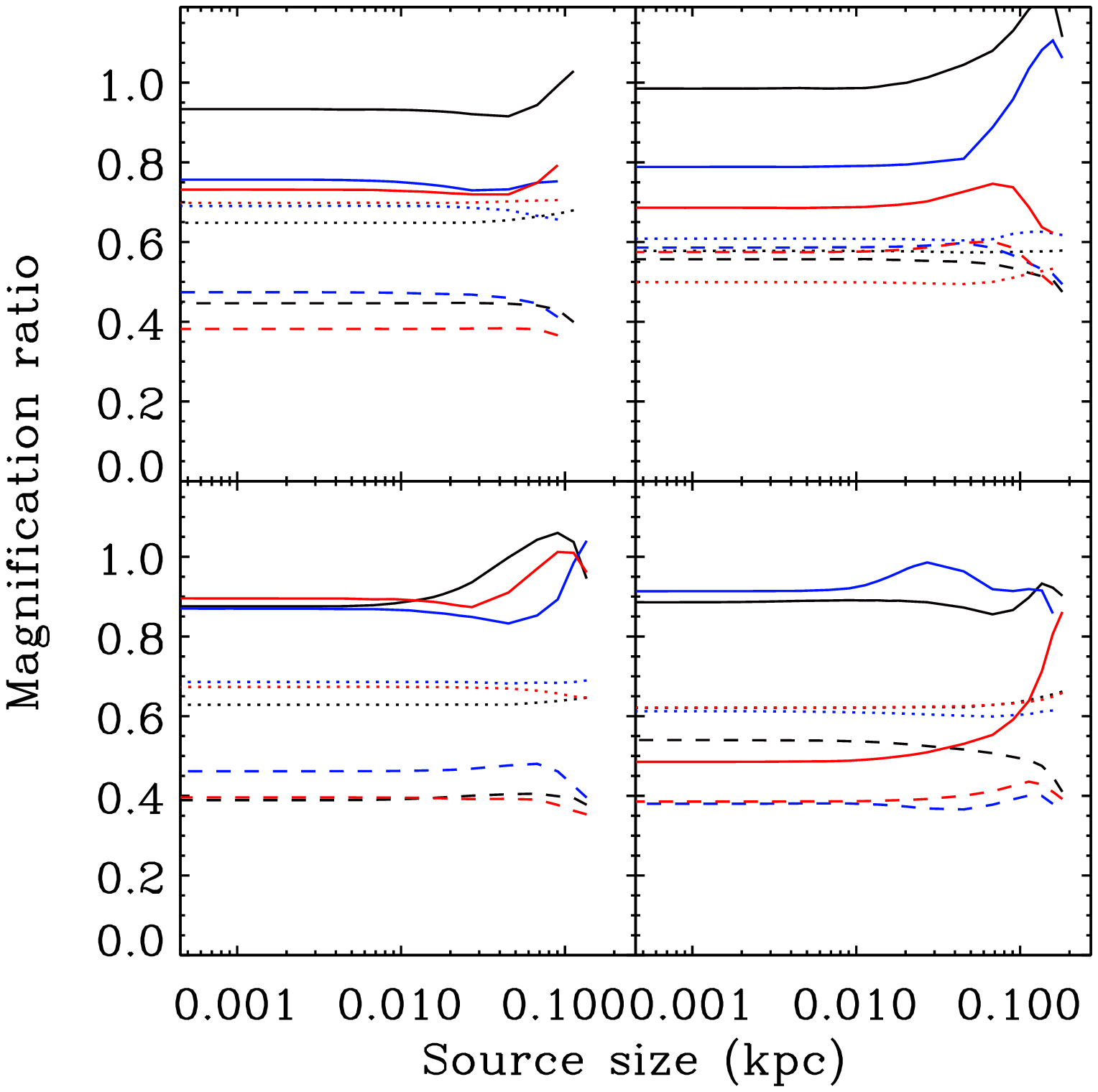}
\caption{\footnotesize The magnification ratios
  of the four bright images as a function of source size for the
  realization of the inclined disk (30$^\circ$) {\it without
    smoothing}.  The solid curve shows the magnification ratios 
  of the close image image pair, the dotted curve shows the
  magnification ratio of the other two images, and the dashed curves
  shows the ratio of the average of each pair.  
Shown here are the results of ray tracing through the original simulated galaxy and
  halo (black) and the results from simulations with random noise
  (blue and red) without smoothing.  The cusp configurations are shown on the
  left and the fold configurations on the right, in the same order as
  in figure~\ref{fig:magratio_pos}.  Much of the
  variations in the ratios seen here are caused by shot noise as can be
seen by comparing this figure to figure~\ref{fig:magratio_fold} where
the shot noise has been removed by smoothing.  Also note that it is
the solid curves that show the largest dependence on source size.} 
\label{fig:magratio_cusp}
\end{center}
\end{figure}

\begin{figure}
\begin{center}
\epsfxsize=15pc
\epsfbox{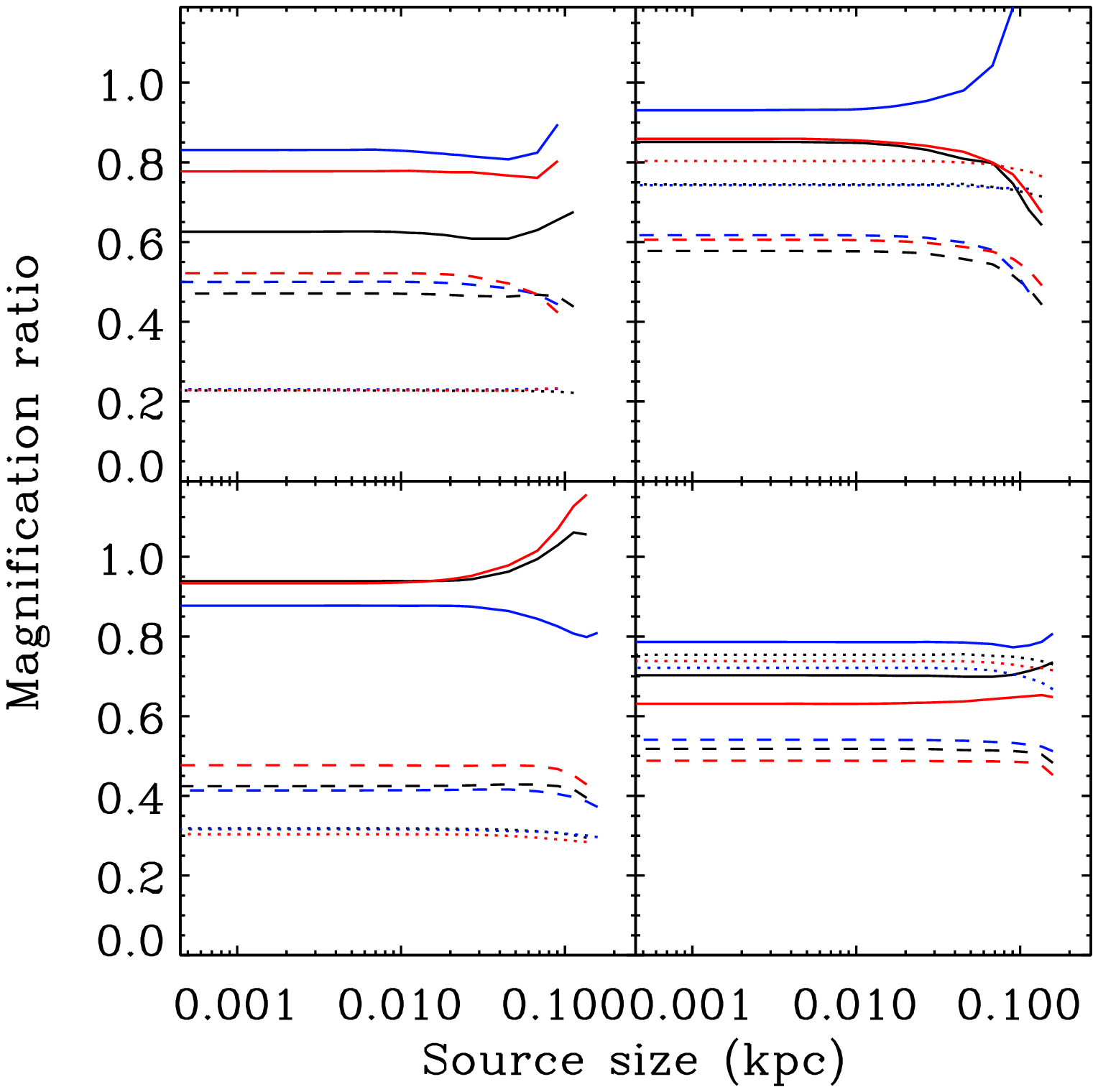}
\epsfxsize=15pc
\epsfbox{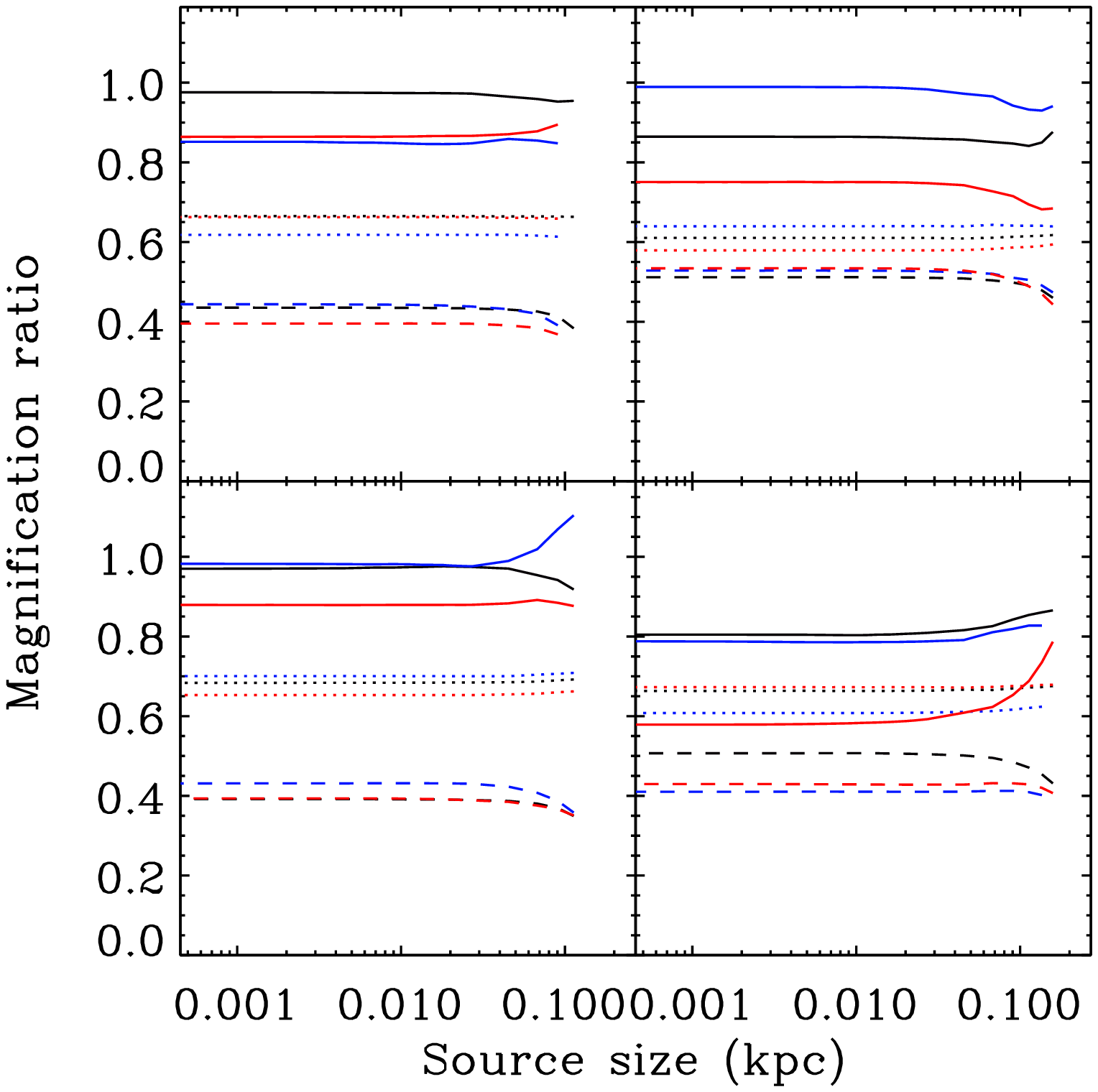}
\caption{\footnotesize The same quantities as in
  figure~\ref{fig:magratio_cusp}, but with the simulated galaxy/halo smoothed
with a Gaussian kernel of width $2\sigma=0.5\kpc$. 
The variations
in the magnification ratios as a function of source size are
significantly smaller than in figure~\ref{fig:magratio_cusp} where no
smoothing was applied.  The variations are largely isolated to the
ratio for the closest pair of images (solid curves). 
}\label{fig:magratio_fold}
\end{center}
\end{figure}

Modeling lenses in search of substructure has proven to be difficult
due to degeneracies in the lens models.  \cite{cirpass2237} used the technique of
spectroscopic gravitational lensing to overcome some of these
degeneracies.  This technique relies on the assumption that without
substructure the magnification ratios will be independent of the
source size (for sizes $\simlt 1 \kpc$).  In this section, we seek to
test this assumption for our simulated lenses in different
image configurations and thus determine
whether a mismatch in the magnification ratios at different
wavelengths can be considered strong evidence for the presence of
substructure.

Our simulated halo has a resolution of $1.68\times10^6\rm
\msun$ so it will not have any substructure less massive
than a $\sim 8\times10^7\msun$, which is at the very limit of the mass range for
the substructure found in \cite{cirpass2237}.  For this reason we do
not expect to be able to represent the kind of structure that could be responsible
for those results. Nonetheless, we can verify the assumption that larger scale
irregularities in the lens will not cause mismatches in the ratios at
different wavelengths.  

We look at the sensitivity of the magnification ratios to source size in three types of image
configurations.  First, the Einstein cross configuration shown
in figure~\ref{fig:magratio_cross} which has four relatively
symmetrically spaced images.
The second configuration comes from placing the source close to a cusp in the
caustic.  This produces three images that form close together and one
isolated image. By placing a source close to each of the four cusps we
produce four versions of this configuration, shown on the left in
figure~\ref{fig:magratio_pos}.  Thirdly, we place the source 
near a fold caustic forming a pair of closely spaced images with a further two
images that are more widely spaced.  Again, by placing a source near
each fold, we have four realizations of this fold caustic configuration
(shown on the right of figure~\ref{fig:magratio_pos}). 

figure~\ref{fig:magratio_cross} shows the results for an Einstein cross
configuration with the disk galaxy inclined at 30$^\circ$.  Based on the
assumptions made in \cite{cirpass2237} we would not expect to see any
significant variation in magnification ratio in this case and there are
none seen.  The magnification ratios being a strong function of source
size is a strong indication that there are structures smaller than the
ones represented in these simulations.

figure~\ref{fig:magratio_cusp} shows the cusp and fold caustic results
with no smoothing and figure \ref{fig:magratio_fold} shows the results 
for the surface densities smoothed with a Gaussian of width
$2\sigma=0.5\kpc$.  
A potential complication with
the interpretation of the differential magnification ratios is that
when the source is large it could overflow the radial caustic region,
especially in cusp and fold caustic configurations.  In this case,
part of the source has four images and part of it has two images.  
This potential problem can be reduced by a judicious choice of
magnification ratios.  The closest two images are usually images of only the
part of the source that is inside the tangential caustic.  The other two
images are of all parts of the source that are within the
radial caustic, usually a superset of the former region.  In theory, the
magnification ratios of this image 
pair should be the least sensitive to source size -- at
least when the source is large.  We single out these ratios in the figures.
For the third ratio, we choose to use the ratio of the average flux
from each pair.  This ratio is potentially affected by the source's
``overflow'' of the tangential caustic region.

  For both the fold and cusp caustic cases, with smoothing
  (figure~\ref{fig:magratio_fold}), we find 
that the magnification ratios are insensitive to source size
while the size is below several hundred pc.  Any variations
are mostly isolated to the ratios between the closest pair of
images.  In general, the further images are separated, the less their
magnification ratio will depend on source size.  This is consistent with
what we had expected -- that the images close to the critical curves,
which have high magnifications, will respond to small variations in the
critical curves introduced by substructure (figure \ref{fig:inc} Panel (d)).
The ratios for the wide separation pairs (dotted curves in
  figure~\ref{fig:magratio_fold}) are effectively 
  independent of source size and thus a large observed dependence can be
considered a trustworthy indicators of smaller scale substructure or
  some other complicating effect.
 
We also see, by comparing figures~\ref{fig:magratio_cusp} and
\ref{fig:magratio_fold}, that the shot noise, or more generally
smaller scale structure, has a significant effect on the magnification ratios
(figure~\ref{fig:magratio_cusp}), but the generic feature that the
ratios are stable for sizes below 10~pc is preserved in all cases and
the variations that are present are largely isolated to the most closely spaced
image pair.  It is also evident from figures~\ref{fig:magratio_cross}
through \ref{fig:magratio_fold} that the values of the magnification
ratios are dependent on the realization of the shot noise.  We interpret
this as the influence of changes in the surface density that are too
large in scale to produce a variation with source size. None the less, they are still
significant enough to change the magnification ratios.

\subsection{The Cusp Caustic Relation}
\label{sec:cusp-relation}

\begin{figure}
\begin{center}
\epsfxsize=40pc
\epsfbox{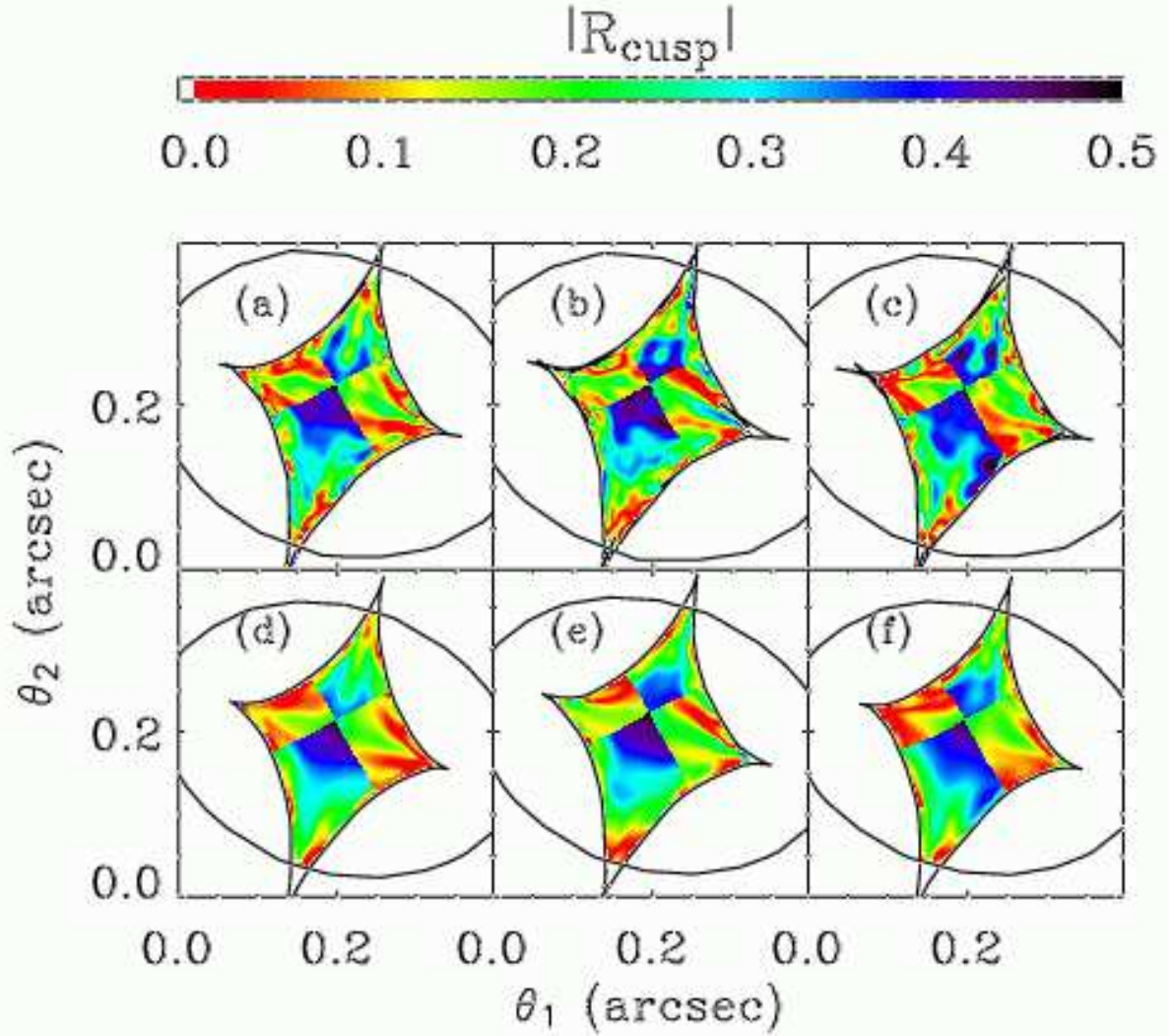}
\caption{\footnotesize Value of the quantity $R_{\rm cusp}$ across the
  four image region of the source plane for each realization of the
  noise using a 60~pc source.  
Panel (a) shows the results for our inclined disk realization, (b) and
(c) show the same quantities with the addition of random shot noise.
Panels (d), (e) and (f) show the corresponding results when the
surface density is smoothed with a Gaussian of width 0.5~kpc.  The
discontinuous jumps are where the three closest images change. 
Not all the regions near the cusps are red as would be expected from the
cusp caustic relation.  There are ``swallow tails'' in the caustics that
appear to be associated with significant violations in the relation.
} 
\label{fig:cusp_ran}
\end{center}
\end{figure}

\begin{figure}
\begin{center}
\epsfxsize=40pc
\epsfbox{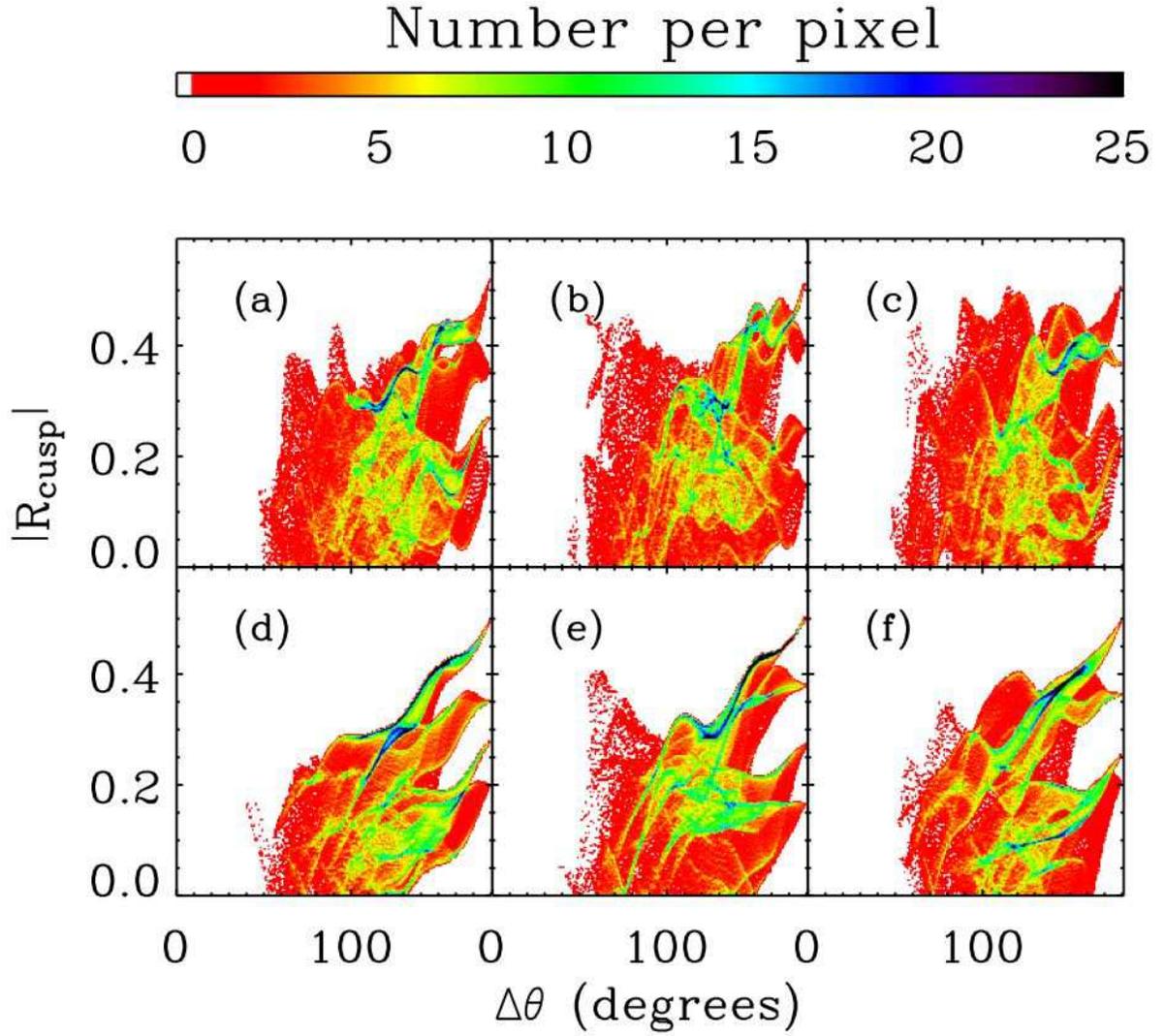}
\caption{\footnotesize Variation in $R\rm _{cusp}$ as a function of
  the angular separation $\Delta\theta$ of the closest three
  images. The results here are for a source with a radius of
  60~pc. Panel (a) shows the results for our inclined disk 
  realization, (b) and (c) show the same quantities with the addition
  of random shot noise as in figure~\ref{fig:cusp_ran}.  Panels (d),
  (e) and (f) show the 
  corresponding results when the surface density is smoothed with a
  Gaussian of width 0.5~kpc.  The color coding is explained in the
  text.}
\label{fig:cusp_scat}
\end{center}
\end{figure}

\begin{figure}
\begin{center}
\epsfxsize=20pc
\epsfysize=25pc
\epsfbox{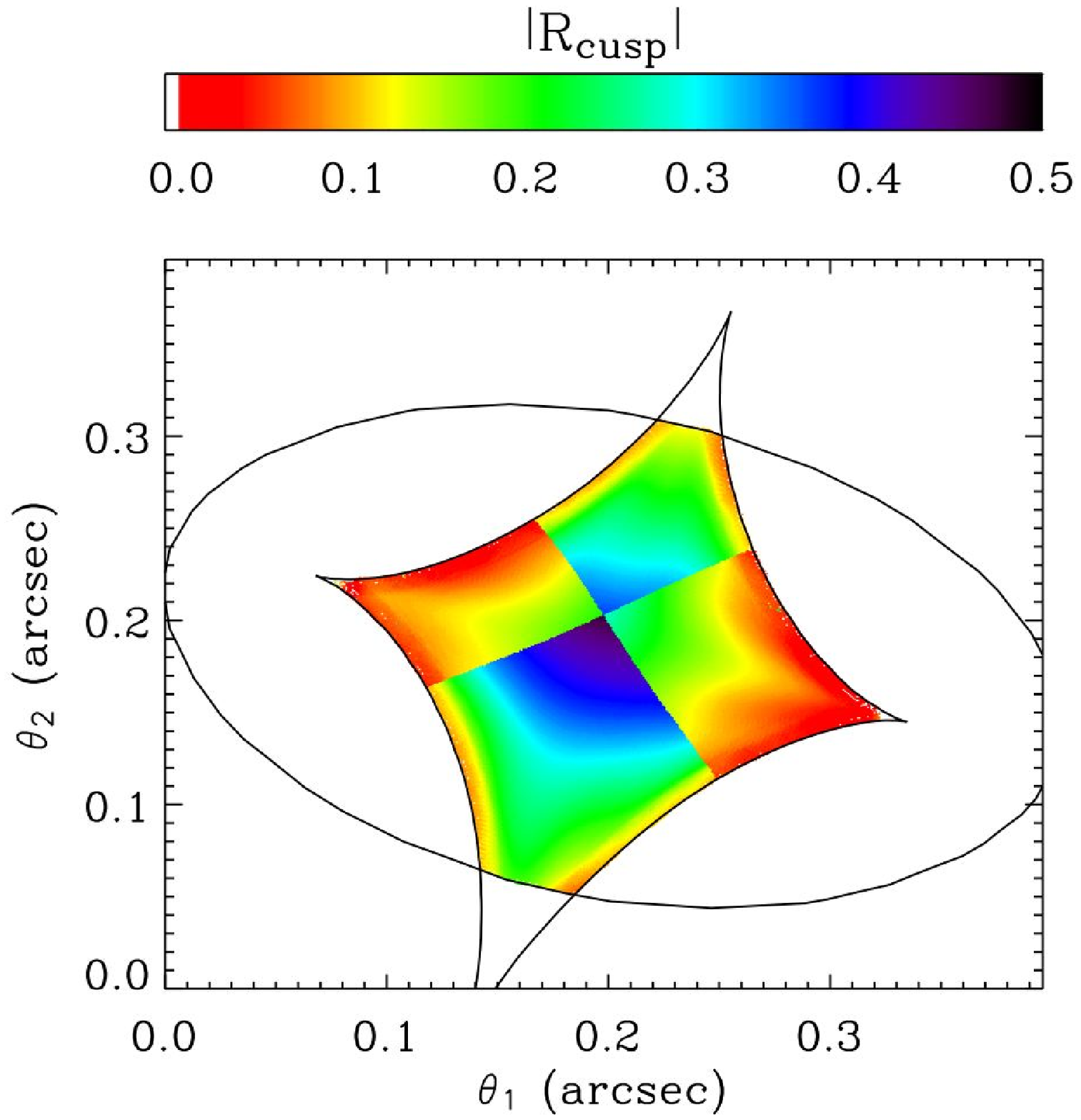}
\epsfxsize=20pc
\epsfbox{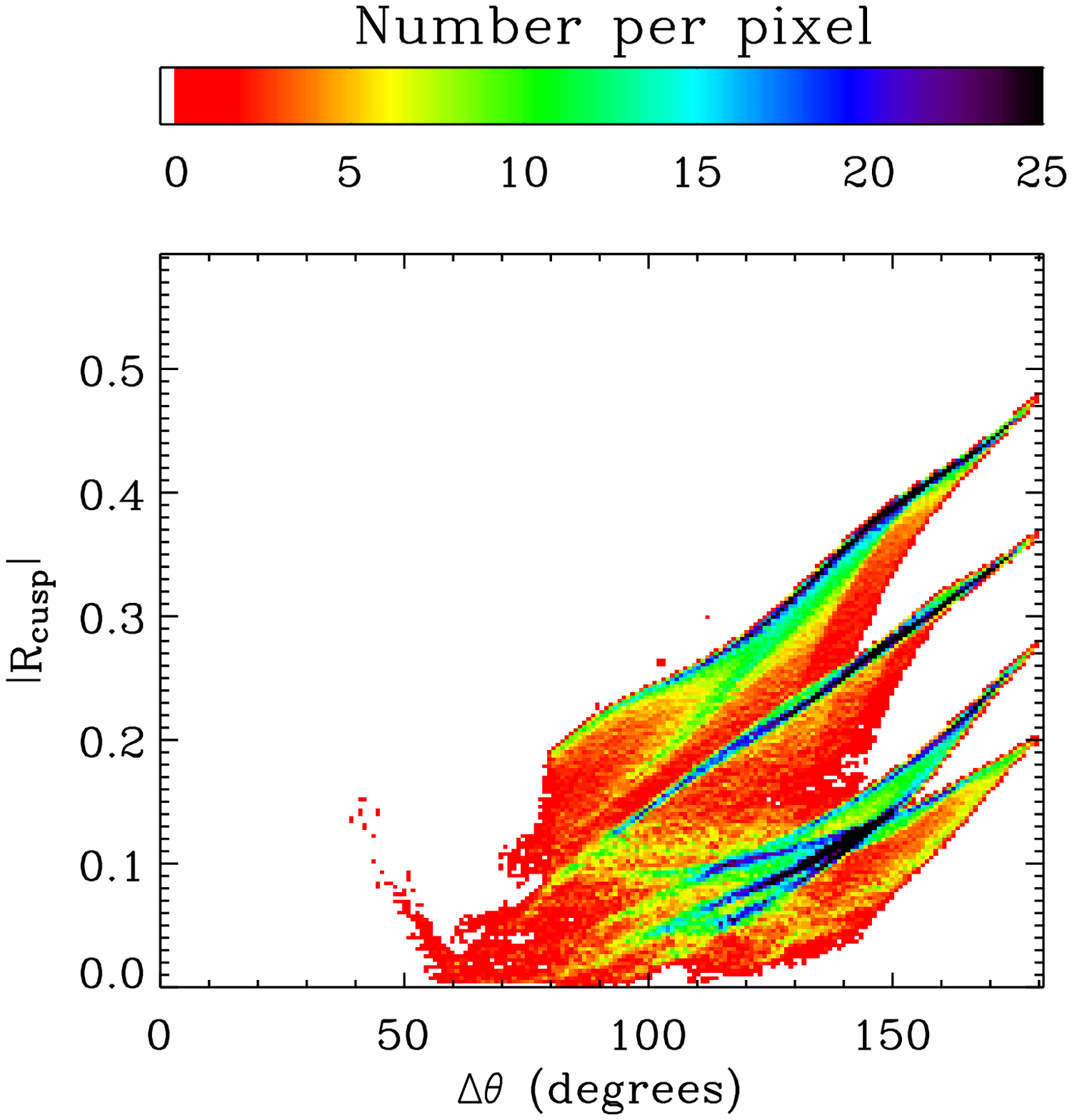}
\caption{\footnotesize The same as figures~\ref{fig:cusp_ran} and
  \ref{fig:cusp_scat} except with a smoothing length of
  $2\sigma=1.0\kpc$. No random noise has been added.  The tangential
  caustic has expanded because the smoothing has reduced the
  concentration of the halo. }
  \label{fig:cusp_smooth}
\end{center}
\end{figure}
Evidence that the cusp caustic relation is violated may provide a way of 
probing the abundance and nature of substructure in lensing galaxies and/or
small-scale structure in intergalactic space. 
All the observed cusp caustic lenses violate this relation at some
level \citep{1998MNRAS.295..587M,KGP2002}.  In order to understand the
significance of such results, it is important to study the ways in
which this relation is affected by structures in the lensing galaxy.  
\cite{KGP2002} has studied the cusp caustic relation in a wide range
of smooth analytic lens models and  \cite{astro-ph/0306238} have looked
at the relation in simulations. In this section, we examine violations
of the cusp caustic relation for our simulated galaxy plus halos and
attempt to establish the number of violations that should be expect in $\Lambda$CDM.

We study the cusp caustic relation by placing a source inside both the
radial and tangential caustics.  Source positions in this region
produce 5 images, with the fifth image being close to the center of
the lens.  For the work in this section we discard this central image
from the analysis routines since it is usually hard to identify due
to its demagnification.  The remaining four images consist of two images
with positive parity and two images with negative parity.

Identifying the images of extended sources is often difficult when their
magnifications are high, as they are in cusp caustic configurations.
The images can be drastically stretched and curved near the outer
(tangential) critical curve.
In such a situation, a single image can be identified as two or more separate
images due to limited grid resolution. In this study, we have decided to "err on the side of
caution" and included only the results for source positions where
we find four outer images with a total parity of zero (+,+,-,-).  As in
the previous section, we have performed consistency checks on the
calculated magnifications with multiple methods.  We find 
that both of these independent methods give consistent results.

For each of the lenses studied, the region of the source plane which
produces four images is divided into a two dimensional grid with a grid
spacing of 10~pc.  This gives us roughly 
70,000 source positions to study.  We place one source of radius 60~pc
at each grid point, producing the four images. We then
identify the three closest images, which we define as the three images
that subtend the smallest opening angle using the peak of the surface
density as the center of the co-ordinate system, denoted
$\Delta\theta$.   These images allow us to measure the cusp caustic parameter,
\begin{equation}
R_{\rm cusp}=\frac{\mu_1+\mu_2+\mu_3}{|\mu_1|+|\mu_2|+|\mu_3|}.
\end{equation}
We also check that the
middle image of the three has the opposite parity to the other
two images.  This process is repeated for all the grid points. 

The cusp caustic relation states that as the three images move close
together, the quantity $R_{\rm cusp}$ should tend to zero
\citep{1986ApJ...310..568B,1992A&A...260....1S,1995A&A...293....1Z,KGP2002}.
Figure~\ref{fig:cusp_ran} shows our results, with panel (a) 
showing our calculations using the disk galaxy inclined at 30$^\circ$
and panels (b) and (c) showing the results for the inclined disk with
the addition of shot noise.  Panels (d), (e) and (f) show results
corresponding to the same realizations as (a), (b) and (c), but smoothed
with a Gaussian kernel of width $2\sigma=0.5\rm kpc$.
Figure~\ref{fig:cusp_smooth} shows the same simulation with a smoothing
length of $2\sigma=1.0\kpc$.

We see a similar variation in $R_{\rm cusp}$ over the source plane as
seen by \cite{astro-ph/0306238} in their elliptical galaxy simulation,
but less variation than in their disk galaxy simulation.  This could be because of
their larger particle masses.  Shot noise does seem to have a larger
influence on their results.  Without smoothing, shot noise has significant effect on the features in our $R_{\rm cusp}$ map.  This
is not surprising. As we have already 
discussed, images with high magnification (close to the critical
curves) are most susceptible to shot noise.  Gaussian smoothing has the
expected result of reducing the variation of $R_{\rm cusp}$ over the caustic,
as well as reducing the discrepancy between the realizations with and
without the addition of shot noise.  Figure~\ref{fig:cusp_smooth}
shows that when the simulation is over-smoothed a regular and
expected pattern in $R_{\rm cusp}$ emerges.
Despite this, there are still significant
regions near the cusps in the $2\sigma=0.5\kpc$ simulation where
$R_{\rm cusp}$ is not small.  The importance of these regions needs to be quantified.

Figures~\ref{fig:cusp_scat} shows $|R_{\rm cusp}|$ as a function of
the angular separation of the closest three images.  The  $|R_{\rm
  cusp}|$, $\Delta\theta$ plane is divided into a 200$\times$200 grid,
forming pixels of $0.9^\circ\times0.003.$  The color scale shows the number
of results that fall in a given pixel, thereby the figure can
be thought of as a 2D histogram, where more likely combinations are green
or blue and the less likely being red.  Panels (a), (b), and (c) show a great deal
of scatter.  figure~\ref{fig:cusp_smooth} shows the same type of plot
only with the surface density smoothed with $2\sigma=1.0\kpc$.  It is necessary to define what is meant by a violation 
of the cusp caustic relation since we would not expect
$R_{\rm cusp}$ to be precisely zero even without substructure.  
\begin{figure}[t]
\begin{center}
\epsfxsize=25pc
\epsfbox{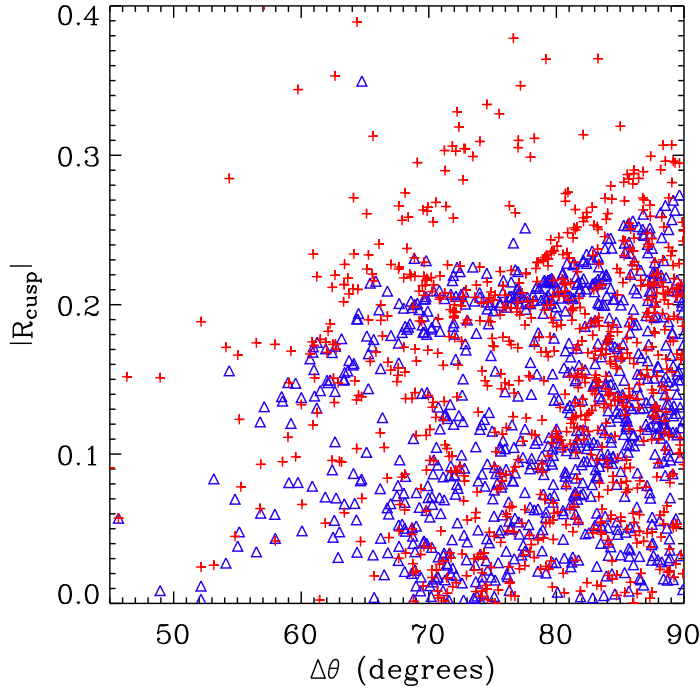}
\caption{\footnotesize Comparing the effect of source size of the cusp caustic relation.  Here we show the  small angle cases that we have classified as cusp cases for source sizes of 60~pc (blue triangles) and 5~pc sources (red crosses).  We see that the smaller sources violate the cusp caustic relation more often.}
\label{fig:cusp_scat_5pc}
\end{center}
\end{figure}

There is an interesting underlying pattern in the histograms on the $R_{\rm
  cusp}$-$\Delta\theta$ plane.  This pattern becomes more defined and
clearer with increased smoothing, being the clearest in figure~\ref{fig:cusp_smooth}.
$R_{\rm cusp}$ is restricted to 4 discrete values
when $\Delta\theta \sim 180^\circ$, each corresponding to the
different quadrants of the asteroid caustic.   This is not surprising
since $\Delta\theta=180^\circ$ corresponds to one point on the source
  plane.  The pattern is a property of the particular lens so its
  details should not be considered universal.

The above results are calculated for sources with
  a radius of 60~pc.  This is large compared to many of the
  sources in real strong lensing systems, and from
  section~\ref{sec:diff-magn-rati} we know that the magnifications in
  the simulation can changed with source sizes below 10~pc.  However,
  repeating the above calculations with a much smaller source size
  would require too much CPU time with our current implementation.
  Small images can be difficult to follow, especially in images with large tangential
  stretch, when the source size drops to roughly 2 orders of magnitude
  below the resolution of the initial grid.  
  To gain an insight into the possible effect of modelling smaller
  sources, instead of using all the source positions we select a
  random sample of $\sim 800$ source points that have been classified as cusp cases
  from the realization with 0.5~kpc of smoothing. $\rm R_{cusp}$ for
  these points is then calculated for a source size of 5~pc.  A
  comparison between the 60~pc and 5~pc sources is shown in
  figure~\ref{fig:cusp_scat_5pc}.  From figure~13 we see that results
  due to image separations of less than 50$^\circ$ can not be trusted
  and are likely to be due to numerical effects.  We therefore
  concentrate our comparison to range $50^\circ < \Delta\theta <
  90^\circ$.  It should also be noted that measuring the properties of
  small sources is a simpler more stable process than dealing with
  larger sources since large sources produce highly extended images
  that can fragment  and hence are susceptible to numerical
  instabilities. 
  It can be seen that there are generally
  more high $\rm R_{cusp}$ cases for the smaller source size, this is not unexpected since the large sources are equivalent to averaging over a number of neighbouring smaller sources, which could lead to greater scatter for the small sources.  The
  significance of these high $\rm R_{cusp}$ cases will be discussed in
  section~\ref{sec:very-brief-comp}. 

\subsection{A Brief Comparison With Observations}
\label{sec:very-brief-comp}

\begin{table}[t]
\begin{center}
\begin{tabular}{l||l|l|c}
&&& {\footnotesize observed} \\
lens & $\Delta\theta$ & $R_{\rm cusp}$  & {\footnotesize band} \\
\hline
B0712+472 & $79.8^\circ$ & $0.26\pm 0.02$\tablenotemark{1,2}~ & radio \\
B2045+265 & $35.3^\circ$ & $0.501\pm 0.035$\tablenotemark{2}~ & radio \\
B1422+231 & $74.9^\circ$ & $0.187\pm 0.006$\tablenotemark{2,3}~~ & radio \\
RXJ1131-1231 & $69.0^\circ$ & $0.355\pm0.015$\tablenotemark{4}~ & optical/IR \\
RXJ0911+0551 & $69.6^\circ$ & $0.192\pm 0.011$\tablenotemark{5}~ & optical/IR 
\end{tabular}
\caption{\footnotesize The image opening angles and cusp caustic
  parameters for the observed cusp caustic lenses.}
\label{tbl:R_observed}
\tablenotetext{1}{\cite{1998MNRAS.296..483J}}
\tablenotetext{2}{\cite{2003ApJ...595..712K}}
\tablenotetext{3}{\cite{2001MNRAS.326.1403P}}
\tablenotetext{4}{\cite{2003A&A...406L..43S}}
\tablenotetext{5}{\cite{KGP2002} \& the CASTLES survey}
\end{center}
\end{table}

Properties of the observed cusp caustic lenses are summarized in
table~\ref{tbl:R_observed}.  All these could be considered violations of
the cusp caustic relation in that $\rm R_{cusp}$ is not consistent
with zero.  But even for a very smooth lens $\rm R_{cusp}$ would not
be exactly zero in all these cases.   The important question is
whether the observed $|\rm R_{cusp}|$ values are generally too large to
agree with the simulations.  

To answer this question we devise a statistical test.  First we define
a cusp caustic lens as one for which $\Delta\theta < 90^\circ$.
The cusp caustic magnification relation should hold asymptotically as
the source gets closer to the cusp or as $\Delta\theta \rightarrow 0$. 
A line can be drawn on the $R_{\rm cusp}$-$\Delta\theta$ plane
starting at $[\Delta\theta,|R_{cusp}|]=[0^\circ,0]$ that divides
``violations'', cases above the line, from ``non-violations'', cases
below the line.  For the line $|R_{cusp}| =
\left(\frac{0.187}{74.9^\circ}\right)\Delta\theta$ all of the observed cases are
classified as violations.  This is the steepest line for which this is
true.  Drawing from the calculations presented in
section~\ref{sec:cusp-relation} we can calculate the fraction of cases
that violate the cusp caustic relation by this definition and the probability that
5 out of 5 cases are violations.  To approximate the
effects of observational noise we add normally distributed random numbers with
a variance of 10\% to the calculated $\rm R_{cusp}$s, a conservative
estimate.  We include 100
realizations of the noise.  Table~\ref{tbl:Rcusp_stats} shows the
fraction of violations for the $30^\circ$ incline disk simulations.
In all cases, even cases without smoothing,  the probability of
getting 5 out of 5 violations is less than 1\%.  The case with the
largest number of violations is the one with a 5~pc source.  But even
in this case the probability of getting 5 out of 5 violations is
estimated at $0.004$.  The observed lenses are in strong disagreement
with the simulations.

The flux ratios of two of the five observed cusp caustic
lenses are available only in optical and infrared wavelengths.  These
magnification ratios could potentially be affected by microlensing by
stars.  The radio fluxes ratios are unlikely to be affected by
microlensing.  Even if we consider only the three lenses with radio
flux ratios the likelihood of getting 3 out of 3 violations is only
0.04 for the 5~pc source case, the one with the most violations. 
The number of observed lenses is clearly 
small, but the disagreement is large.  This indicates that at
least the substructure that is resolved in these simulations is not enough
to cause the observed anomalies.

\cite{KGP2002} have considered these same lenses and concluded that
all but B1422+231 violate the relation by a different definition based
on analytic lens models.  This is a very crude method for comparing
the simulations to the data. 
It is possible to make a more restrictive comparison by incorporating more
information about individual lenses.  

We have considered only our simulated disk galaxy where
the number of violations to the cusp caustic relation is expected to
be larger than for elliptical galaxies.  Most lenses are ellipticals
so we expect that the discrepancy between the observations and the
simulations is even larger than reported here.

In addition to the observed cusp caustic cases, the results of
section~\ref{sec:diff-magn-rati} show that the 
assumption of spectroscopic gravitational lensing are sound.  The
discrepancy between the radio, mid-infrared and narrow line
magnification ratios of Q2237+0305 \citep{cirpass2237} cannot be
accounted for by $\Lambda$CDM halos because there are too few with
masses $\simlt 10^7\msun$.  This result has yet to be confirmed with
other lenses or adequately explained.

\begin{table}[t]
\begin{center}
\begin{tabular}{l||c|c|c|c}
& No smoothing & 2$\sigma$ = 0.5~kpc & 2$\sigma$ = 0.5~kpc & 2$\sigma$ = 1.0~kpc \\
source size &  60~pc & 60~pc & 5~pc & 60~pc \\
\hline
\hline 
S $30^\circ$ spiral & 0.31 & 0.23 & 0.34 & 0.00 \\
S random 1          & 0.25 & 0.18 & - & - \\
S random 2          & 0.31 & 0.24 & - & - \\
\end{tabular}
\caption{\footnotesize The ratio of simulated cases that violate the cusp caustic
  relationship ($|R_{\rm cusp}| >
  \left(\frac{0.187}{74.9^\circ}\right) \Delta\theta$) to the total 
number of simulations with $\Delta\theta<90^\circ$.  Random noise has been
  added at account for observational uncertainties at the level of 10\%.}
\label{tbl:Rcusp_stats}
\end{center}
\end{table}

\section{Conclusions and Discussion}
\label{sec:concl--disc}

Galactic sized gravitational lenses have been simulated by combining a 
cosmological N-body simulation with models for the baryonic component.  
The lensing properties have been calculated through a combination of
ray-shooting and adaptive mesh refinement with the goal of studying
the effects of the substructure.

There are two main conclusions: First, the image magnifications in the Einstein
cross configuration are very weak functions of source size when it is
below $\sim 1\kpc$.    In the fold and cusp caustic cases we find the
same thing for the magnification ratios of widely separated images.
This confirms the belief that spectroscopic gravitational lensing can be
used to detect small-scale structure \citep{MM02,cirpass2237}. Second, if
there is no substructure below $\sim 0.5\kpc$, we expect that the cusp
caustic relation (by the definition $|R_{\rm cusp}| >
  \left(\frac{0.187}{74.9^\circ}\right) \Delta\theta$) would hold for
  the majority of the lenses.  None of the five observed cusp caustic
  lenses satisfy this requirement.  The probability of this happening
  is less than 1\% in all the simulations we investigated.

It might be that all the halo substructure produced in the $\Lambda$CDM is
not enough to account for the cusp caustic violations.  The amount of
substructure in the simulations used here ($\simgt 8\times10^7\msun$)
is not enough, and not much mass in
compact enough substructures is expected below our resolution.  This argument was made by
\cite{2004ApJ...604L...5M} and depends on how much
substructure is destroyed in the inner regions of the lens halo.  It
should be possible to see explicitly if smaller scale 
substructures contribute to the lensing using future simulations and the
techniques developed here.  \cite{Metcalf04} recently argued that
the observed cusp caustic anomalies can be accounted for by
intergalactic halos within the $\Lambda$CDM model.  Distinguishing
between these contributions will be important. 

We have tried to estimate the uncertainties in the lensing properties
caused by the limited resolution of the simulation by introducing random
realizations of the estimated noise to the surface density.  It appears 
that we may have over estimated
the actual shot noise in the simulations, due to the fact that the simulations
with noise added tend to have more irregular critical curves and
$R_{\rm cusp}$ distributions than the original simulation.  This
makes our estimates of the uncertainties in the lensing properties
more conservative.  \cite{astro-ph/0306238} in a similar study estimated
the noise by displacing individual particles at random.  
Shot noise had a larger affect on their simulations probably because of
their larger particle masses.  We do not see as many swallow tail caustics and
violations of the cusp caustic relation as \cite{astro-ph/0306238} did.
We believe that this is because of their limited resolution and the
different radial profile of their lenses.

We have used only one simulated dark matter halo in this study so small
number statistics could be an issue in interpreting the results.
It could be that a bias has been introduced because the simulated dark
matter halo used here was preselected to be relatively isolated and
dynamically relaxed.  A random sample of halos might have more irregular
structure with remnants of recent mergers within them.  This could produce
more anomalous lensing behavior than what is seen here.  For example, there
is a prominent substructure in the halo that is within 2'' of
the center of the galaxy in the projection shown in
figure~\ref{fig:inc}.  There could be several of these in a less relaxed halo.

The limiting factors in this study has been the resolution of the
cosmological simulation and the absence of baryons in the substructure.  Because of the limited force and mass
resolution, we expect that the central cores of substructure may be
destroyed.  It is interesting to note that lensing through the
realizations without smoothing does not produce significantly more
violations of the cusp caustic magnification relation, not enough to be in agreement with observations.
Since these unsmoothed cases contain close to the maximum granularity
  in the mass range $10^{4}$ to $10^{7} \msun$, this indicates that
  the discrepancy between observation and prediction may not be solved
  simply by performing higher resolution simulations.  
The next generation of cosmological N-body simulations should resolve
substructure up to 2 orders of magnitude lower.  Baryons in the
small dark matter halos, if allowed to cool, will condense into the
centers of substructures making them more concentrated which makes
them less susceptible to tidal disruption and more efficient lenses.
This will only be important for relatively high mass substructures
however ($\simgt 10^8\msun$) because the condensed baryons would
necessarily produce stars and there are not enough dwarf galaxies with
masses $\simlt 10^8\msun$ to account for the number of dark matter
subhalos in the simulations
\citep{1999ApJ...524L..19M,1999ApJ...522...82K}.  Hydrodynamic
simulations with star formation have not yet reached the level of
resolution needed for this kind of lensing study.  This is a topic
that needs more investigation.  
With improvements in the simulations, the kind of study presented in
this paper will be able to conclusively compare the
lensing data with the predictions of $\Lambda$CDM.

\acknowledgments
\footnotesize
We thank B. Moore for providing us with the simulated halo.  AA also
thanks T. Saini and L. King for helpful discussions. We would like to
thank the anonymous referee for constructive comments.  Financial
support for this work was provided by NASA through Hubble Fellowship 
grant HF-01154.01-A awarded by the Space Telescope Science Institute,
which is operated by the Association of Universities for Research 
in Astronomy, Inc., for NASA, under contract NAS 5-26555

\bibliography{mybib}
\bibliographystyle{apj_jss}

\end{document}